\documentclass[preprint,aip,pof]{revtex4-2}
\usepackage[dvipdfmx]{graphicx}
\usepackage{siunitx} 
\usepackage{dcolumn}
\usepackage{bm}
\usepackage{color}
\usepackage{array}
\usepackage{tabu}
\usepackage{amsmath,amssymb,stmaryrd}
\usepackage{amsfonts}
\usepackage{subfigure}
\usepackage{subfigmat}
\usepackage{pbox}
\usepackage{xfrac}
\usepackage{tabularx}
\usepackage{multirow}
\usepackage[percent]{overpic}
\usepackage{float}
\usepackage[linesnumbered,ruled,vlined]{algorithm2e}
\usepackage{algorithmic}
\newcolumntype{M}[1]{>{\centering\arraybackslash}m{#1}}
\makeatletter
\def\BState{\State\hskip-\ALG@thistlm}
\makeatother
\def\drawline#1#2{\raise 2.0pt\vbox{\hrule width #1pt height #2pt}}

\usepackage{bm}

\newcommand{\fg}[1]{{{\color{black}#1}}}
\graphicspath{{./figures/}}

\begin{document}


\title{Experimental velocity data estimation for imperfect particle images using machine learning} 

\author{Masaki Morimoto}
\email[]{masaki.morimoto@kflab.jp}
\affiliation{Department of Mechanical Engineering, Keio University, Yokohama 223-8522, Japan}

\author{Kai Fukami}
\email[]{kfukami1@g.ucla.edu}
\affiliation{Department of Mechanical and Aerospace Engineering, University of California, Los Angeles, CA 90095, USA}
\affiliation{Department of Mechanical Engineering, Keio University, Yokohama 223-8522, Japan}

\author{Koji Fukagata}
\email[]{fukagata@mech.keio.ac.jp}
\affiliation{Department of Mechanical Engineering, Keio University, Yokohama 223-8522, Japan}

\date{\today}

\begin{abstract}
\baselineskip 18pt
We propose a method using supervised machine learning to estimate velocity fields from particle images having missing regions due to experimental {limitations.}
As a first example, a velocity field around a square cylinder at Reynolds number of ${\rm Re}_D=300$ is considered.
To train machine learning models, we utilize artificial particle images (APIs) as the input data, which mimic the images of the particle image velocimetry (PIV).
The output data are the velocity fields, and the correct answers for them are given by a direct numerical simulation (DNS).
We examine two types of the input data: APIs without missing regions (i.e., full APIs) and APIs with missing regions (lacked APIs).
The missing regions in the lacked APIs are assumed following the exact experimental situation in our wind tunnel setup. 
The velocity fields estimated from both full and lacked APIs are in great agreement with the reference DNS data in terms of various statistical assessments.
We further apply these machine learned models trained with the DNS data to experimental particle images so that their applicability to the exact experimental situation can be investigated.
The velocity fields estimated by the machine learned models contain approximately 40 folds denser data than that with the conventional cross-correlation method.
This finding suggests that we may be able to obtain finer and hidden structures of the flow field which cannot be resolved with the conventional cross-correlation method.
We also find that even the complex flow structures are hidden due to the alignment of two square cylinders, the machine learned model is able to estimate the field in the missing region reasonably well.
The present results indicate a great potential of the proposed machine learning based method as a new data reconstruction method for PIV.

\end{abstract}


\maketitle 
\section{Introduction}
\label{intro}

High-resolution fluid data in both space and time are essential to advance our understanding on complex fluid flow phenomena.
To meet this requirement, a vast range of measurement techniques have been proposed and improved to date, such as hot-wire anemometry, Schlieren photography, laser Doppler velocimetry, and particle image velocimetry (PIV).
Among these, PIV has widely been used due to its capability of obtaining a velocity field in a two dimensional plane (i.e., planar PIV) or even in a three dimensional volume (i.e., tomographic PIV).
For PIV, the cross-correlation method~\cite{adrian2005} has been used as the major technique to estimate velocity fields from particle images.
The mean velocity of the particles in a finite sized region called an interrogation window is determined by searching the region which has the maximum correlation of intensity distribution among two consecutive images in time. 
Another method is the optical flow method proposed by Horn and Schunck~\cite{HS1981}.
In contrast to the concept of pattern matching in the cross-correlation method, the optical flow method is able to estimate velocity vectors by a spatio-temporal gradient of intensity distribution and to obtain higher resolution data than the cross-correlation method.
However, one of the big issues here is that the error would increase significantly when the velocity magnitude increases because the time differential term is included in the optical flow formulation.
Therefore, proper experimental equipments, i.e., a high-spec high-speed camera and a high-power laser emitter, are indispensable to apply the optical flow method in order to obtain particle images in a short stride of time.

\begin{figure}[t]
    \centering
    \includegraphics[width=0.8\textwidth]{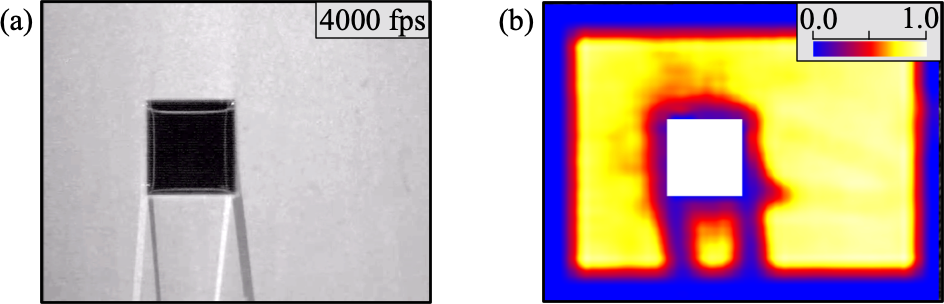}
    \caption{(a) A typical particle image for a flow around a square cylinder made of transparent acrylic resin, taken by the authors using the affordable PIV system described in Appendix A. The air flows from left to right. Illumination is done by a laser sheet from the top. The halation light and shadow regions significantly affect the velocity estimation. (b) Distribution of correlation coefficient. Correlation is quite low in the regions where the illumination is not appropriate in addition to the outer edges of the image. The image size here is $640\times 480$ pixels, and the interrogation window is $32\times 32$ pixels.}
    \label{fig:1-1}
\end{figure}

Not only for the methods introduced above, estimation of velocity fields in PIV is directly related to the intensity distribution of particle images.
Therefore, the experimental images need to be as noiseless and appropriately brightened as possible. 
However, we often encounter cases where we cannot avoid regions with insufficient particle images, especially in the cases with immersed objects.
For instance, the laser sheet is irregularly reflected on the surface of an object.
Also, an object having edges causes extremely white or dark regions due to halation and shadow.
An example of such a particle image around a square cylinder, taken by the authors using an affordable high-speed camera based PIV system (see, Appendix A for details), is shown in Fig.~\ref{fig:1-1}(a).
Although the square cylinder is made of transparent acrylic resin, we can observe white and dark regions below the cylinder due to its corners.
Also, it can be seen in Fig.~\ref{fig:1-1}(b) that the correlation coefficient {is significantly low} in the regions around and under the cylinder {in addition to the outer edges of the image}, which suggests that velocity estimation {using the cross-correlation method} is poor in those regions.

To overcome such issues, several techniques have been proposed to date.
One of the well-known techniques is to paint an object surface in flat black so that the laser light can be absorbed~\cite{PMDC2013}.
This can reduce the effect of irregular reflection and improve the velocity estimation.
However, because the reflection cannot be prevented perfectly, it is still difficult to acquire the particle images near the surface.
Moreover, since the model is painted, information in the shaded region is totally lost.
The other candidate is a combination with laser induced fluorescence (LIF), as has been used for PIV of liquid-gas two-phase flow to overcome the intense light reflection due to gas bubbles~\cite{Fujiwara2004}.
{In this case,} tracer particles are colored by fluorescent paint that emits light at a different frequency from that of the irradiated laser sheet, and by observing only the light emitted from particles the influence of reflection and halation can be mitigated almost perfectly.
Although the combination with LIF performs very well, its use for single-phase gas flow experiments is burdensome because the fluorescent paint, e.g., rhodamine B, usually includes carcinogenetic material~\cite{Rhodamine2007}.
Therefore, a new method which can accurately estimate a velocity field without the aforementioned issues is eagerly desired.

As a proof of concept of surrogates for the aforementioned conventional methods, various methods have been proposed for flow data estimation and reconstruction from limited measurements.
These efforts can be seen not only in experimental but also in computational fluid dynamics due to the high demand for high-resolution flow data.
Gappy proper orthogonal decomposition (Gappy POD) proposed by Everson et al.~\cite{ES1995} has been considered as one of the candidates to reconstruct a flow field from incomplete data.
Bui-Thanh et al.~\cite{BDW2004} applied the Gappy POD to a flow around an airfoil and showed that the flow field can be successfully reconstructed from snapshots with 30\% data missing.
However, the Gappy POD assumes that a missing portion differs in different snapshots so that it cannot be applied to the situations where the missing region, for example the shadow region in Fig.~\ref{fig:1-1}, is fixed.
Use of the Kalman filter~\cite{CHBH2006} and the linear stochastic estimation~\cite{SH2017} was also investigated for state estimation of turbulent channel flow.
More recently, applications of machine learning have also emerged to take into account nonlinearities in the schemes~\cite{BEF2019,BNK2020,FFT2020,guastoni2019prediction,SGASV2019,MFRFT2020,FNF2020,AEreview2020,Gentech2020}.
Capability of machine learning based super resolution, which reconstructs high-resolution flow field from its low-resolution counterpart, was demonstrated by Fukami et al.~\cite{FFT2019,FFT2019TSFP} for turbulent flows: the proposed model can substantially recover the energy spectra in the higher wavenumber range of two-dimensional decaying turbulence.
The concept of super resolution has recently been extended to not only computational~\cite{OSM2019,LTHL2020,FFT2020b,GANSRkim2021,Gao2021,LTHL2020,guemes2021coarse} but also experimental data sets~\cite{DHLK2019,wang2020predicting}.
\fg{
This idea can also be extended to a global field reconstruction task from sparse sensor measurements~\cite{erichson2020,FukamiVoronoi,SGHK2021}.
These studies assisted with machine learning open a new path of data-driven modeling in fluid flow analyses such as turbulence modeling~\cite{DIX2019,Duraisamy2021,gamahara2017searching,LKT2016,maulik2018data,wang2018investigations,yuan2020deconvolutional,xie2020artificial,yao2020modeling,subel2021data,ren2021priori,ling2015evaluation,zhu2019machine,yang2020improving,yin2020feature,jiang2021interpretable,xie2019modeling,wu2018physics,beetham2020formulating,xie2020modeling,maulik2017neural,maulik2019subgrid,park2021toward,milani2021turbulent,beetham2021sparse}, flow control~\cite{BHT2020,raibaudo2020machine,tang2020robust,ren2021applying,han2020active,ren2020adaptive,li2020machine,zheng2021active,rabault2019accelerating,novati2019controlled,rabault2019artificial,zhou2020artificial,paris2021robust,park2020machine}, reduced-order modeling~\cite{TBDRCMSGTU2017,THBSDBDY2020,pawar2019deep,pawar2020data,maulik2021reduced,agostini2020exploration,renganathan2020machine,eivazi2020deep,pawar2020long,bukka2021assessment,leask2021modal,wu2021reduced,peng2020unsteady,gundersen2021semi,kong2021deep,LW2019,lee2019data}, and data reconstruction~\cite{li2021efficient,sekar2019fast,li2019inversion,guemes2019sensing,giannopoulos2020data,kashefi2021point,ren2020lower,peng2020time,nair2020leveraging,kim2020prediction}.
}

Of particular interest from the recent trends in reconstructing and estimating flow data using machine learning is its application to the PIV data processing.
{To the best of our knowledge, the first application of machine learning to PIV was demonstrated by Teo et al.~\cite{Teo1991} who utilized Fuzzy adaptive resonance theory (Fuzzy ART)~\cite{ART1990}, a multi-layer perceptron consisted with an input layer and a classification layer, to automatically pair the particles of two time consecutive images.}
Chen et al.~\cite{CYC1998} proposed a multi-layer perceptron based estimation method for PIV of a uniform flow and reported that the estimated mean velocity showed reasonable agreement with the result of conventional PIV.
Rabault et al. \cite{RKJ2017} utilized a convolutional neural network (CNN) for the first time to estimate velocity fields from synthetic particle images. 
Moreover, Cai et al.~\cite{CZXG2019} have recently used a machine learning model called FlowNetS architecture~\cite{DFI2015} to estimate velocity fields of a cylinder wake, a backward facing step flow, and isotropic homogeneous turbulence from synthetic particle images and demonstrated that it can estimate higher resolution data than the conventional PIV.
They have also applied the proposed method to experimental particle images of turbulent boundary layer, and the results imply its great potential.
Similarly, Grayver et al.~\cite{GN2020} utilized a supervised machine learning model to predict a velocity field from streak images of particles on both synthetic and experimental data.
Gim et al.~\cite{GJSKK2020} have applied a shallow neural network to detect three-dimensional particle locations for real-time particle detection in three-dimensional particle tracking velocimetry.

\fg{Despite these recent trends, there are still only few studies focusing on applications of machine learning to experimental data.
Moreover, none of them handles cases with an object immersed in a flow which makes a fixed missing region in PIV.}
This paper aims at estimating complete velocity fields from experimental particle images with a fixed missing region due to laser reflection etc. using a supervised machine learning method.
\fg{Since it is a common issue in PIV, it is crucial to investigate the applicability of machine learning-based methods to such noisy measurements, for propelling practical applications of machine learning-based methods for experimental situations.}
An autoencoder-type CNN, which is robust for translation or rotation of images, is considered for the present study.
We utilize artificial particle images (APIs) generated with a direct numerical simulation (DNS) to train the machine learning model and apply the trained model to experimental images.
The present paper is organized as follows.
In Sec.~\ref{sec:methods}, we introduce the overview, the training data, and the machine learning model of the present study.
The results of estimation for APIs and experimental data are presented and discussed in Sec.~\ref{sec:results}.
At last, concluding remarks are offered in Sec.~\ref{sec:conclusion} with some outlooks on experimental data estimation using data-oriented methods.

\section{Methods}
\label{sec:methods}
\subsection{Overview of the present study}
\label{sec:overview}

\begin{figure}
\centering
\includegraphics[width=\textwidth]{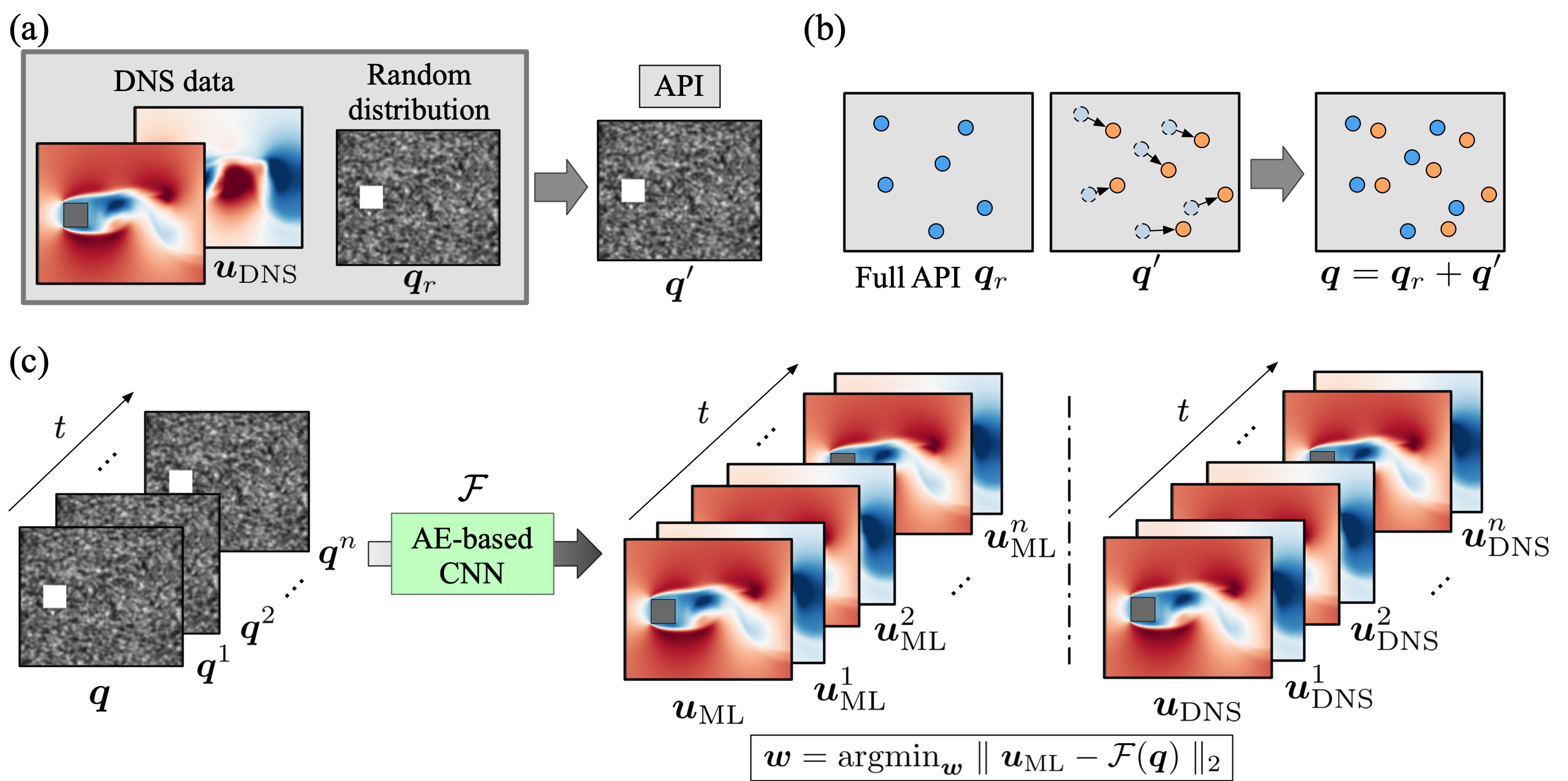}
\caption{Overview of the machine learning based experimental flow data estimation. 
(a) Preparation of training data set. 
The artificial particle image (API) at a certain time instant ${\bm q}_r$ is generated by randomly distributing particles.
The API with a time increment ${\bm q}^\prime$ can be obtained using the velocity data ${\bm u}_{\rm DNS}$.
(b) Schematic of the API used for the input to the machine learning model, $\bm{q} = \bm{q}_r+\bm{q}^\prime$.
(c) Training of machine learning model $\cal{F}$ for full data.}
\label{fig:flowchart}
\end{figure}

Let us introduce an overview of the present study in Fig.~\ref{fig:flowchart}.
For a training process of supervised machine learning, data sets comprising of pairs of an input {$\bm{q}$} and the solution {$\bm{u}$} must be prepared.
Then, the model $\cal F$ is trained such that ${\bm{u}}\approx{\cal F}({\bm{q}})$.
In this study, we aim to establish a machine learning model $\cal F$ to output the velocity fields ${\bm u}=\{u,v\}$ from input particle images {$\bm{q}$.}
Although our final objective is to apply this model to experimental data, pairs of particle images and the corresponding velocity fields obtained by the conventional particle image velocimetry (PIV) data processing method, e.g., the cross-correlation method, are not appropriate as the training data in terms of reliability due to the reflection of a laser light as mentioned above.
To prepare the clean training data, instead of the experimental data with missing regions, we use synthetic particle images generated with the aid of direct numerical simulation (DNS).
Details for numerical {setup} of the present DNS can be found in Appendix B.

As the first step for preparing {the} training data set, we generate {an artificial} particle image {(API) at a certain time instant} $\bm{q}_r$, as shown in Fig.~\ref{fig:flowchart}(a).
The particles are randomly distributed here. 
Details on the generation of APIs will be offered {in Sec.~\ref{sec:API}.}
We utilize the DNS data to calculate the location{s} of {these} particles {to obtain the API with a small time increment} $\bm{q}^\prime$.
{Then, the images at two time steps $\bm{q}_r$ and $\bm{q}^\prime$ are summed up to} obtain the API $\bm{q}$ as illustrated in Fig.~\ref{fig:flowchart}(b).
This procedure enables us to emulate the experimental particle images {obtained by a classical double exposure or streaky particle images taken during a relatively long} exposure time~\cite{GN2020}. 
The APIs at different time instants are generated likewise.

As shown in Fig.~\ref{fig:flowchart}(c), the next step is the construction of the machine learning model $\cal F$ for the full APIs $\bm{q}$.
The model $\cal F$ attempts to predict velocity fields ${\bm u}=\{{u},{v}\}$ from the APIs $\bm{q}$.
Hence, the objective of training process for the model $\cal F$ is to obtain optimized weights $\bm w$ so as to estimate the velocity fields from the APIs by minimizing the difference between estimated velocity fields and reference DNS data used to generate APIs, such that
\begin{equation}
    \bm{w}=\mathrm{argmin}_{\bm{w}}||\bm{u}_\mathrm{DNS}-\mathcal{F}(\bm{q};{\bm w})||_2.
    \label{eq:minimizeW}
\end{equation}
In the present study, we choose $L_2$ error norm as the loss function.

Subsequently, we also construct the other machine learning model $\cal{\widehat{F}}$ to estimate velocity fields $\bm u$ from the lacked APIs $\widehat{\bm{q}}$, which is analogous to Fig.~\ref{fig:flowchart}(c) but has a data missing region corresponding to the low correlation region in Fig.~\ref{fig:1-1}(b).
The lacked APIs {$\widehat{\bm{q}}$ are} generated {by applying a mask to the} full APIs $\bm{q}$, {and they} include {a} lacked portion around a square cylinder.
Similar to the case of full APIs, the training process for the model $\cal{\widehat{F}}$ can be formulated as
\begin{equation}
    \widehat{\bm{w}}=\mathrm{argmin}_{\widehat{\bm{w}}}||\bm{u}_\mathrm{DNS}-\widehat{\mathcal{F}}(\widehat{\bm{q}};{\widehat{\bm w}})||_2,
    \label{eq:minimizeW_forhat}
\end{equation}
where $\widehat{\bm{w}}$ represents weights of the machine learning model $\cal{\widehat{F}}$.
At last, we evaluate the applicability of the machine learned model $\cal{\widehat{F}}$ to experimental particle images.

\subsection{Artificial particle image}
\label{sec:API}

\begin{figure}
\centering
\includegraphics[width=0.6\textwidth]{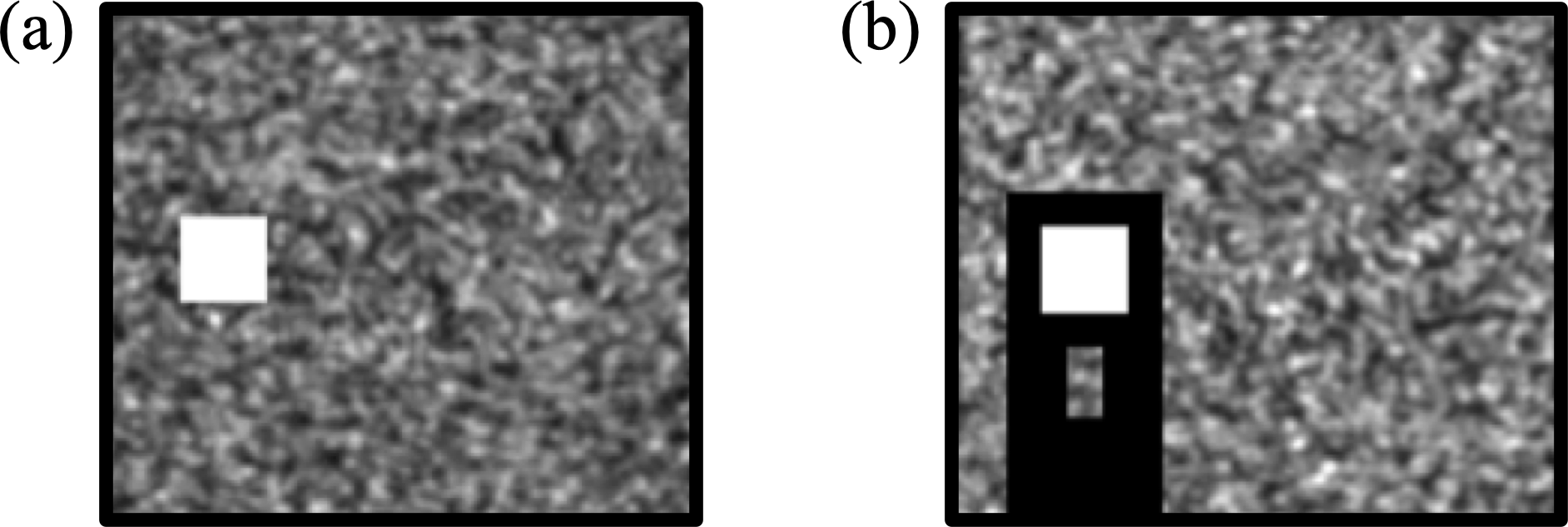}
\caption{Examples of artificial particle images (APIs) around a square cylinder: (a) full API ${\bm q}={\bm q}_r+{\bm q}^\prime$; (b) lacked API $\widehat{\bm q}=\widehat{\bm q}_r+\widehat{\bm{q}^\prime}$. 
Lacked APIs $\widehat{\bm q}$ are prepared by applying a mask to the full APIs.}
\label{fig:API_sample}
\end{figure}

As introduced in Sec.~\ref{sec:overview}, we use the artificial particle images (APIs) as the input for training the machine learning models.
Figure~\ref{fig:API_sample} shows samples of full and lacked APIs.
The APIs in the present study are generated based on the formulation proposed by Okamoto et al.~\cite{ONSK2000}.
The intensity $I\in[0,1]$ at location $(x,y)$ is defined as
\begin{eqnarray}
    I(x,y)&=&\sum_{i=1}^{N}I_{0,i}\exp\left(-\frac{(x-x_{p,i})^2+(y-y_{p,i})^2}{(d_p/2)^2}\right),
    \label{eq:I}\\
    I_{0,i}&=&0.06\exp\left(-\frac{z^2_{p,i}}{\sigma_l^2}\right),
    \label{eq:I_0}\\ \nonumber
\end{eqnarray}
where $(x_p, y_p, z_p)$ is the location of a particle, $d_p$ is the particle diameter, $\sigma_l$ is the laser sheet thickness, and $N$ is the number of particles, respectively.
In this study, we set $d_p=3$ pixels and $\sigma_l=2$ pixels by referencing the experimental images of our experimental setup.
The intensity around each particle is defined with Gaussian distribution, whose 
maximum value should be tuned so as to roughly mimic the experimental images.
In this study, we set it at $0.06$ following our preliminary tests.
{For the lacked APIs, we set a lacked region so as to mask the low correlation regions in our experimental situation and substitute zero there.}

\begin{table}
\centering
\caption{Intensity fitting of APIs to the experimental image.}
\label{tb:API_hist}
\begin{tabular}{cccc}
\hline\noalign{\smallskip}
& \multicolumn{1}{c}{\multirow{2}{*}{Experimental image}} & \multicolumn{2}{c}{Artificial particle image}\\
& \multicolumn{1}{c}{}  & w/o processing & w/ processing\\ 
\noalign{\smallskip}\hline\noalign{\smallskip}
Average  & 34.8    & 53.0 (52.6\%)    & 34.5 (0.85\%)\\
Variance & 8.24    & 8.60 (4.37\%)    & 9.04 (9.71\%)\\
Skewness & 2.34    & 2.53 (8.15\%)    & 2.28 (2.66\%)\\
Flatness & 6.44    & 7.74 (20.1\%)    & 6.04 (6.23\%)\\
\noalign{\smallskip}\hline
\end{tabular}
\end{table}
\begin{figure}
    \centering
    \includegraphics[width=\textwidth]{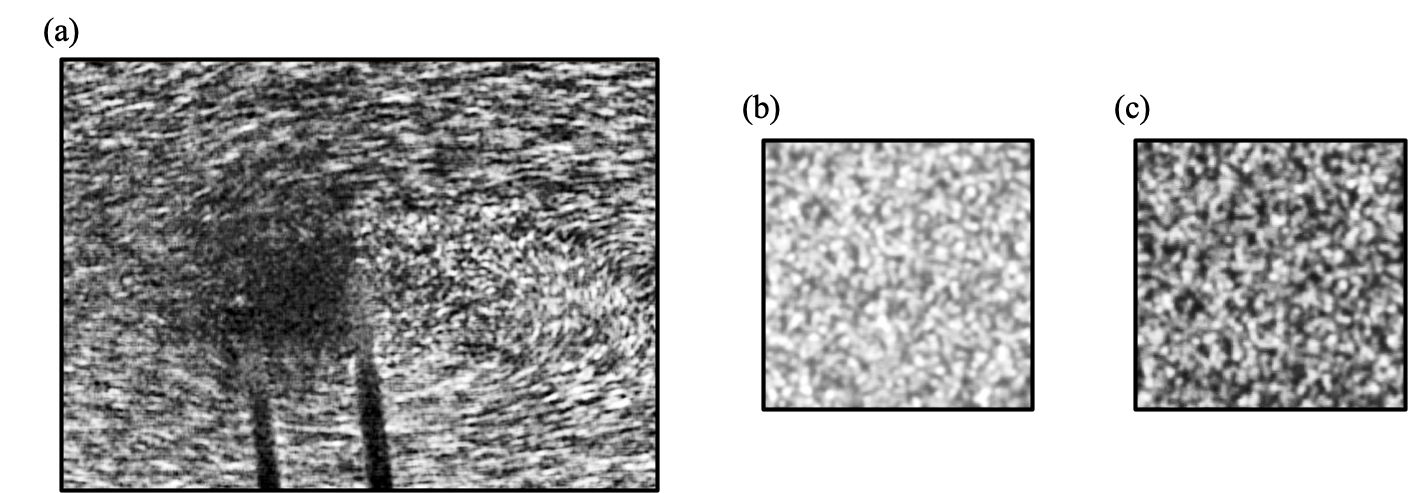}
    \caption{An example of intensity fitting for APIs. (a) Experimental particle image, (b) API without and (c) with post-processing.
    Note that the present experimental image is rather similar to that of the particle streak velocimetry~\cite{GN2020} than that of the classical double exposure.
    }
    \label{fig:API_histmatch}
\end{figure}

Because the APIs must emulate experimental particle images, we further adjust the intensity distribution of the APIs to that of the experimental images.
For this adjustment, we utilize {\tt{imhistmatch}} in MATLAB function, which can account {for the first four statistical moments (i.e.,} average, variance, skewness factor, and flatness factor) of the intensity distribution.
The values of {those moments} of the APIs with and without post-processing are summarized in Table~\ref{tb:API_hist}.
As the result of the histogram fitting, the intensity histogram of the post-processed API shows great agreement with {that} of the experimental image.
Although {the} variance is not improved, the post-processed API shows better results than without processing as a whole, since the average value is drastically improved.
We also present in Fig.~\ref{fig:API_histmatch} the samples of experimental image and fitted APIs.
It is found that the adjusted API can emulate the experimental image better than the original image.
For readers' information, we use a PIV laser G1000 (Kato Koken) and a high-speed camera K5-USB 8GB (Kato Koken) with AF Nikkor $85\ \si{mm}$ F/1.6D (Nikon) lens to capture the particle images of flow fields.
Further details and additional information on our PIV setup can be found in Appendix~A.
Note that {the particle streaks in the present image are relatively long due to the limitation of our system; however, in general, the streak length may be optimized}
through an optimization of parameters~\cite{CYC1998} within the capability of PIV system that will be used.

\subsection{Machine learning model}

\begin{figure}
    \centering
    \includegraphics[width=\textwidth]{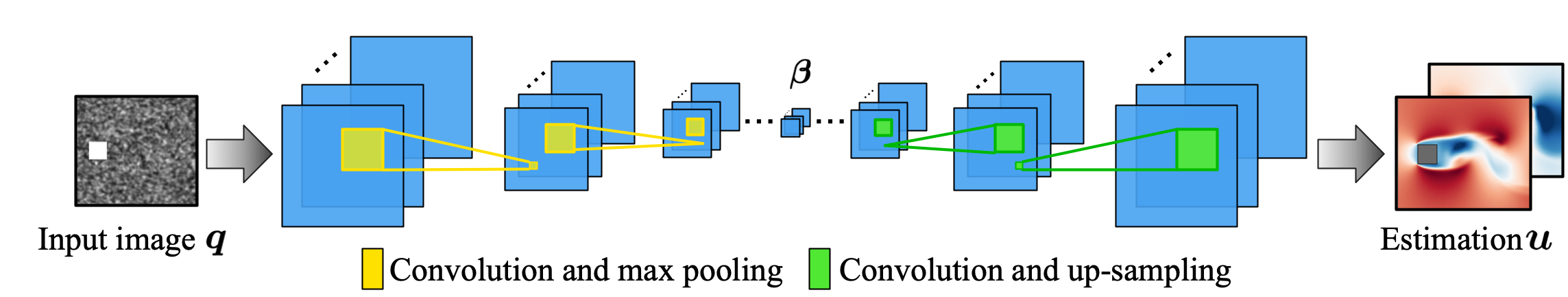}
    \caption{Autoencoder-based convolutional neural network used in the present study.}
    \label{fig:CNN}
\end{figure}
\begin{table}
    \centering
    \caption{Details of the proposed autoencoder-based convolutional neural network.}
    \vspace{3mm}
    \label{tab:str_of_CNNAE}
    \begin{tabular}{ccccc}
    \hline\noalign{\smallskip}
         Layer & Filter size & $\#$ of filters & Data size & Activation function\\
         \noalign{\smallskip}\hline\noalign{\smallskip}
         Input & - & - & (120,140,1) & - \\
         1st Conv2D & (5,5) & 32 & (120,140,32) & ReLU\\
         2nd Conv2D & (5,5) & 32 & (120,140,32) & ReLU\\
         1st Max pooling & - & - & (24,28,32) & -\\
         3rd Conv2D & (5,5) & 32 & (24,28,32) & ReLU\\
         4th Conv2D & (5,5) & 32 & (24,28,32) & ReLU\\
         2nd Max pooling & - & - & (12,14,32) & -\\
         5th Conv2D & (5,5) & 32 & (12,14,32) & ReLU\\
         6th Conv2D & (5,5) & 32 & (12,14,32) & ReLU\\
         3rd Max pooling & - & - & (6,7,32)   & -\\
         7th Conv2D & (3,3) & 16 & (6,7,16) & ReLU\\
         8th Conv2D & (3,3) & 16 & (6,7,16) & ReLU\\
         1st Upsampling & - & - & (12,14,16) & -\\
         9th Conv2D & (5,5) & 32 & (12,14,32) & ReLU\\
         10th Conv2D & (5,5) & 32 & (12,14,32) & ReLU\\
         2nd Upsampling & - & - & (24,28,32) & -\\
         11th Conv2D & (5,5) & 32 & (24,28,32) & ReLU\\
         12th Conv2D & (5,5) & 32 & (24,28,32) & ReLU\\
         3rd Upsampling & - & - & (120,140,32) & -\\
         13th Conv2D & (5,5) & 32 & (120,140,32) & ReLU\\
         14th Conv2D & (5,5) & 2 & (120,140,2) & Linear\\
         \noalign{\smallskip}\hline
    \end{tabular}
    \label{tb:network_structure}
\end{table}

In the present study, a model structure similar to a convolutional neural network (CNN) based autoencoder is used to estimate velocity field $\bm u$ from the input particle image ${\bm q}$.
CNNs~\cite{LBBH1998} have often been utilized in the field of image recognition, and recently, use of CNNs has also been propagated in the fluid dynamics community~\cite{FNKF2019,SP2019,MFF2019,HFMF2020a,HLC2019,LY2019b,nakamura2020extension,fukami2020sparse,morimoto2021convolutional,matsuo2021supervised,nakamura2021comparison,MFNMNF2021}.

The CNN is mainly composed of convolutional layers and pooling layers.
The convolutional layer extracts key features of input data by filtering operation,
\begin{equation}
    c_{ijm}^{(l)}=\psi\left(\sum^{K-1}_{k=0}\sum^{H-1}_{p=0}\sum^{H-1}_{q=0}c_{i+p-C,j+q-C,k}^{(l-1)}h_{pqkm}+b_{m}\right),
\end{equation}
where {$C={\rm floor}(H/2)$}, $c_{ijm}^{(l-1)}$ and $c_{ijm}^{(l)}$ {are} the input and output data at layer $l$, $h_{pqkm}$ denotes a filter of the size of $\left(H\times H\times K\right)$ and $b_m$ is a bias.
The output from the filtering operation is eventually multiplied by an activation function $\psi$.
Using the nonlinear activation function here, a machine learning model can take into account nonlinearity in its structure.
In the pooling layer, representative values are downsampled by pooling operations, e.g., maximum values (max pooling) or average values (average pooling).
It is widely known that by incorporating the pooling operations CNN models can acquire robustness against the variance of input data due to decrease of spatial sensitivity~\cite{LBBH1998}.
Moreover, denoising effect can also be expected by several pooling operations, which is crucial for dealing with experimental images. 

\begin{table}
\centering
\caption{Hyper parameters used in the present machine learning model.}
  \begin{tabular}{cccc}
  \hline\noalign{\smallskip}
    Parameter & Value & Parameter & Value \\
    \noalign{\smallskip}\hline\noalign{\smallskip}
    Batch size & 10 & Time interval of data & 0.25 \\
    Optimizer for network & Adam & Percentage of training data & 70\% \\
    Learning rate of Adam & 0.001 & Learning rate decay of Adam & 0  \\
    $\beta_1$ of Adam & 0.9 & $\beta_2$ of Adam & 0.999 \\
    Number of epochs & 3000 & Early stopping patience & 20 \\
    \noalign{\smallskip}\hline
  \end{tabular}
  \label{tab:param}
\end{table}

One of the recent uses of CNNs is as an autoencoder~\cite{HS2006}, which is trained to output the same data as the input via a low dimensional bottleneck called the latent space. 
Although the output of the present study is not the same as the input as mentioned above, we here borrow the model structure of the CNN autoencoder.
Let us present in Fig.~\ref{fig:CNN} the proposed machine learning model for experimental data estimation.
The details of the proposed model are summarized in Table~\ref{tab:str_of_CNNAE}.
As can be seen in both Fig.~\ref{fig:CNN} and Table~\ref{tab:str_of_CNNAE}, the input vector ${\bm q}$ is mapped into a low dimensional latent space $\bm \beta$ with max pooling operations before the dimension is recovered at the output layer.
As mentioned above, lower spatial sensitivity and robustness against noise can be acquired using the pooling operations.
As the pooling operation downsamples the representative values, e.g., max and average values, it lower{s} the spatial sensitivity and generalizes the model. 
This feature is especially crucial for the present problem setting, since the location of each particle varies randomly in actual experimental images.
Also, the denoising effect can be expected through the pooling operations.
Hence, the CNN autoencoder-like model which has the pooling procedure is suitable for our problem setting since we will apply a model $\widehat{\cal F}$ trained with lacked data to experimental data.
We have checked that the present CNN autoencoder-like model outperforms a regular CNN model without the pooling and upsampling operations in our preliminary test.
As the activation function, we use a rectified linear unit (ReLU), which has widely been utilized because of its stability in updating weights of a training process~\cite{NH2010}.

For training of the present machine learning model, we use the Adam optimizer~\cite{kingma2014} and early stopping criteria~\cite{prechelt1998} to avoid overfitting.
The details of the hyperparameters are summarized in Table~\ref{tab:param}.
Note in passing that the accuracy of the present model can be improved utilizing some theoretical optimization methods, e.g., hyperopt~\cite{BYC2013}, Bayesian optimization~\cite{BCF2009,maulik2019time,LW2019}, and random search~\cite{LW2019}, although we do not consider here.
In the present study, baseline machine learning models of both single and staggered square cylinders are trained using 10000 $x-y$ cross sectional snapshots, which correspond to 500 time steps per twenty positions in the $z$ direction.
The preparation for training data set is detailed in Appendix B.
The dependence of the estimation accuracy on the number of the snapshots $n_{\rm snapshot}$ will be investigated later.

\fg{We emphasize that a size of the image does not 
significantly
affect the present ML-based method since the CNN used in the present study considers a local feature of the image with filter operation.
Namely, once the ML model is established, the computational cost required to obtain the velocity field, i.e., the forward calculation of Eq.~(5), is only proportional to the number of pixels of the image.
Moreover, it should be also possible to treat the input image and perform the estimation in a local manner~\cite{2021wallmodel}.
}

\section{Results and discussion}
\label{sec:results}
\subsection{Example 1: Wake around a square cylinder}

\begin{figure}
    \centering
    \includegraphics[width=\textwidth]{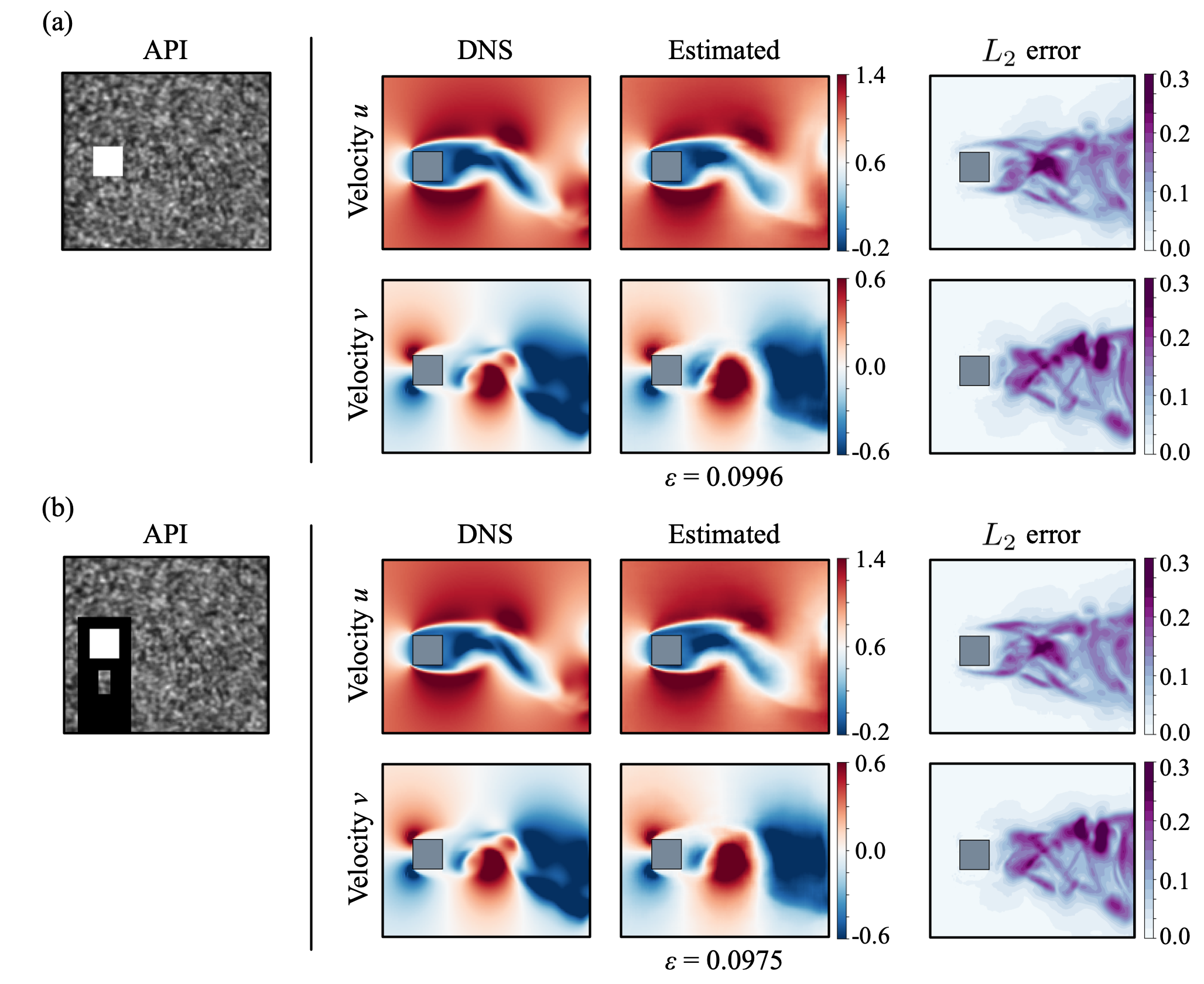}
    \caption{Velocity fields estimated from artificial particle images {(APIs)} of flow around a square cylinder{: (a) full APIs $\bm{q}$; (b) lacked APIs $\widehat{\bm{q}}$.}}
    \label{fig:1SC_API}
\end{figure}

Let us first present in Fig.~\ref{fig:1SC_API} the estimation using the {full and lacked} artificial particle images {(APIs),} $\bm{q}$ and $\widehat{\bm{q}}${, respectively}.
As shown here, the estimated fields with both attributes show great agreement with reference direct numerical simulation (DNS) data.
The $L_2$ error norm $\epsilon=||{\bm u}_{\rm DNS}-{\bm u}_{\rm ML}||_2/||{\bm u}_{\rm DNS}||_2$ for both full and lacked APIs are approximately $10\%$, which indicates that the reconstruction ability is not influenced by the presence of missing regions in the input API.
Note that the distribution of local $L_2$ error norm of velocities are concentrated in the vortex shedding region rather than the missing region where the flow is almost steady in this particular demonstration.

\begin{figure}
    \centering
    \includegraphics[width=\textwidth]{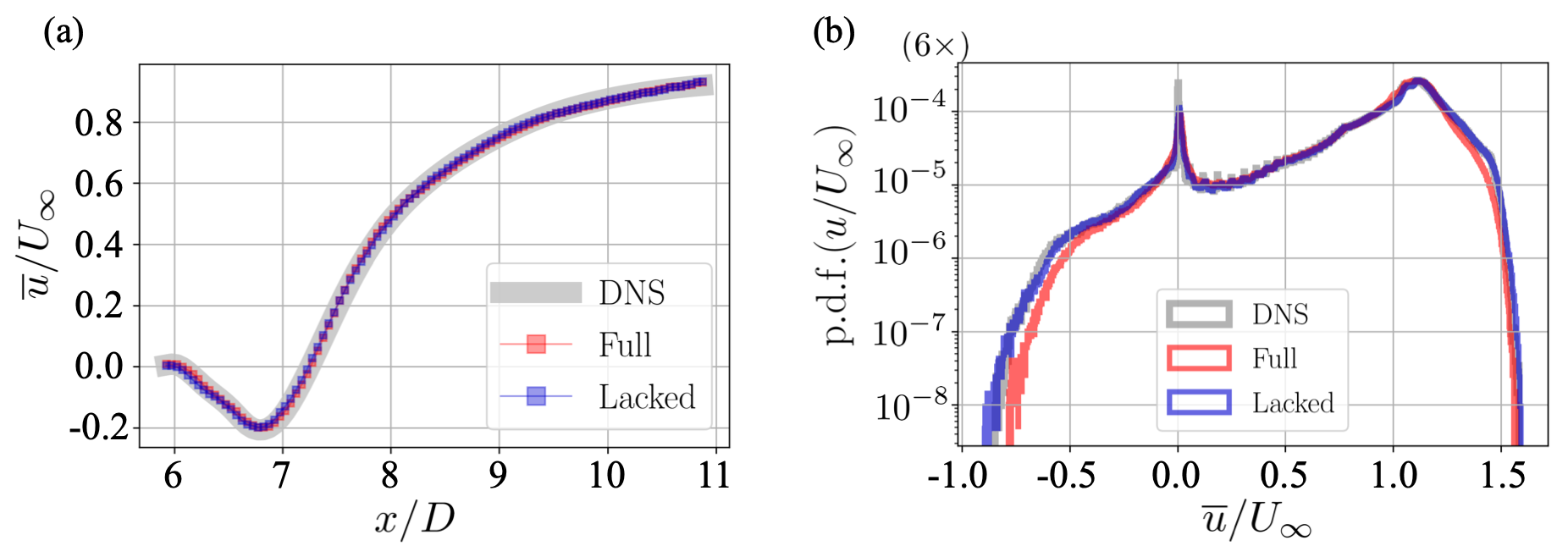}
    \caption{(a) Mean centerline velocity profile and (b) probability density function (PDF) of streamwise velocity estimated with full and lacked APIs of a flow around a square cylinder.}
    \label{fig:1SC_API_umean_PDF}
\end{figure}

For the statistical evaluation, the mean streamwise velocity profile at $y=0$ (i.e., the centerline velocity) and the probability density function (PDF) of the streamwise velocity are shown in Fig.~\ref{fig:1SC_API_umean_PDF}.
The centerline velocities of the fields estimated from the full and lacked APIs show almost perfect agreement with that of DNS.
It also indicates that presence of the missing region has no influence on the estimation ability for this particular demonstration.
The same trends can also be seen in the PDF, as shown in Fig.~\ref{fig:1SC_API_umean_PDF}(b).

\begin{figure}
    \centering
    \includegraphics[width=\textwidth]{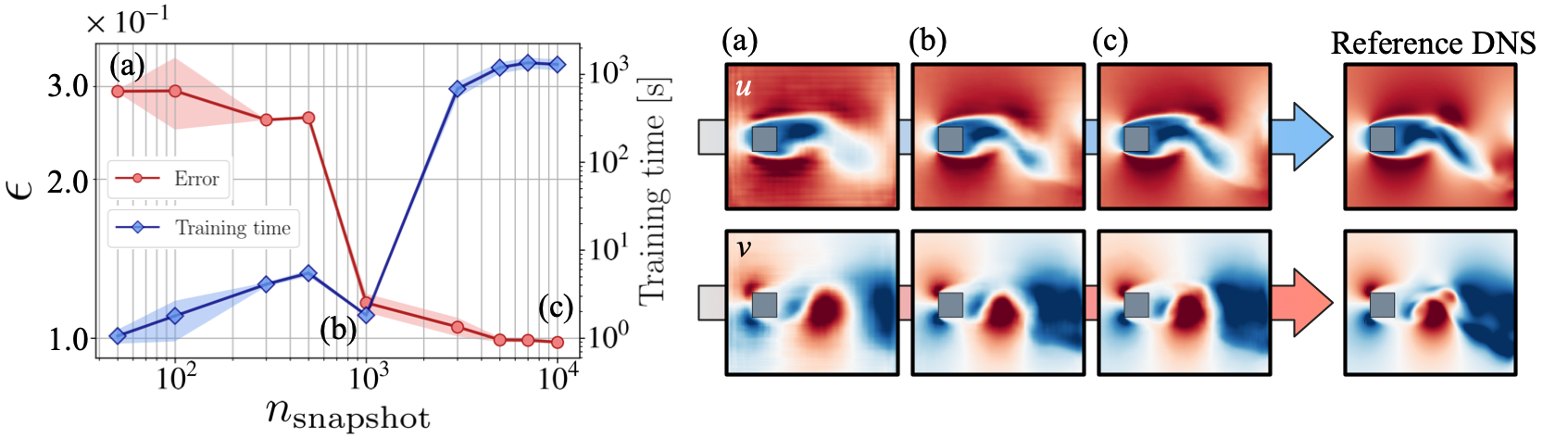}
    \caption{Dependence on the number of snapshots and training time with a single cylinder. \fg{Example velocity fields estimated by models trained with (a)~$n_{\rm snapshot}=50$, (b)~$n_{\rm snapshot}=1000$, and (c)~$n_{\rm snapshot}=10000$ are shown.}}
    \label{fig:1SC_dep_of_NoS}
\end{figure}

Let us examine the dependence of the $L_2$ error norm on the number of snapshots used for training, as presented in Fig.~\ref{fig:1SC_dep_of_NoS}.
Note again that the results {presented} above are obtained by the machine learned model trained with $n_{\rm snapshot}=10000$.
The errors of the estimated fields are relatively high in the case of lower $n_{\rm snapshot}$.
This trend can be seen with $n_{\rm snapshot}=50$ in Fig.~\ref{fig:1SC_dep_of_NoS}(a).
The estimated fields of both attributes are blurry and not well matched with the reference DNS. 
But, noteworthy here is that it can still estimate the large scale motion of the flow from the API with as little as $n_{\rm snapshot}=50$.
The error of the estimated field suddenly decreases around at $n_{\rm snapshot}=1000$ and converges approximately towards $\epsilon=1\times 10^{-1}$ with $n_{\rm snapshot}=5000$ to $10000$.
The sudden expansion of training time at $n_{\rm snapshot}=1000$ is likely because the machine learning model can successfully reduce the loss at $n_{\rm snapshot}=1000$ in this particular case, which causes the delay of convergence in the $L_2$ error. 
Of course, whether the nonlinear trend in a number of snapshots can be seen or not highly depends on problem settings and machine learning models. 
The similar behavior regarding to the error convergence was also observed in our other study~\cite{FFT2020b}.
These findings suggest that although the training time here is relatively short (approximately 20 minutes on the NVIDIA TESLA V100 graphics processing unit (GPU) even in the case with largest number of $n_{\rm snapshot}$), users should care the trade-off relationship between the computational burden and the accuracy of the model.

\begin{figure}
    \centering
    \includegraphics[width=\textwidth]{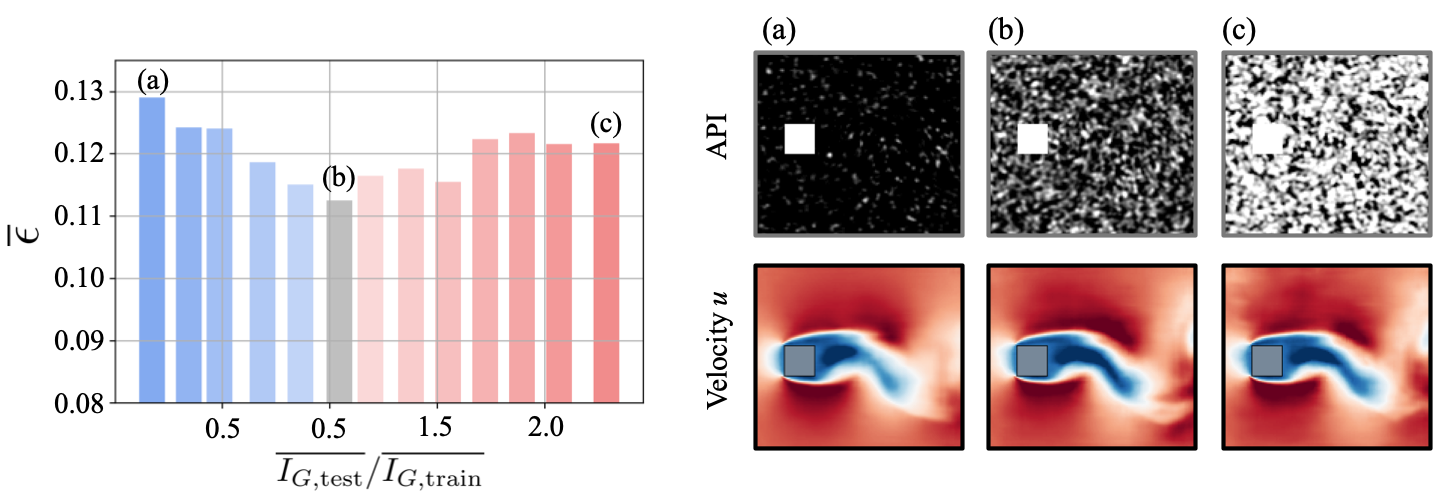}
    \caption{Robustness against image intensities of the test data with a single square cylinder. \fg{Example APIs with the intensity of 
    (a) $I_{G,{\rm test}}/I_{G,{\rm train}}=0.2$, (b) $I_{G,{\rm test}}/I_{G,{\rm train}}=1.0$, and (c) $I_{G,{\rm test}}/I_{G,{\rm train}}=2.4$ are shown with the estimated fields.}}
    \label{fig:intensity}
\end{figure}

Next, we check the applicability and robustness to experimental situations.
For the discussions above, the input APIs are generated with a single intensity $I_{G,{\rm train}}$.
Since the intensity of experimental images varies on each experimental setup, machine learning models are required to be robust to the intensity variance unless the amount and kinds of training data are increased.
Here, let us assess the robustness against the intensity of test images, as presented in Fig.~\ref{fig:intensity}.
The intensity of each image $I_G$ is defined as an average of intensity for each pixel of images.
The estimation using the test images with the same intensity as the training data, i.e., $\overline{I_{G,{\rm test}}}/\overline{I_{G,{\rm train}}}=1$, results in the lowest mean $L_2$ error norm, as expected.
As the test images become darker or brighter than the training images, the $L_2$ error norm increases, but the maximum difference in the $L_2$ error norm is less than $2\%$, as can be seen in Fig.~\ref{fig:intensity}, which suggests that the present machine learned model is robust against intensity variations in the test data. 

\begin{figure}
    \centering
    \includegraphics[width=\textwidth]{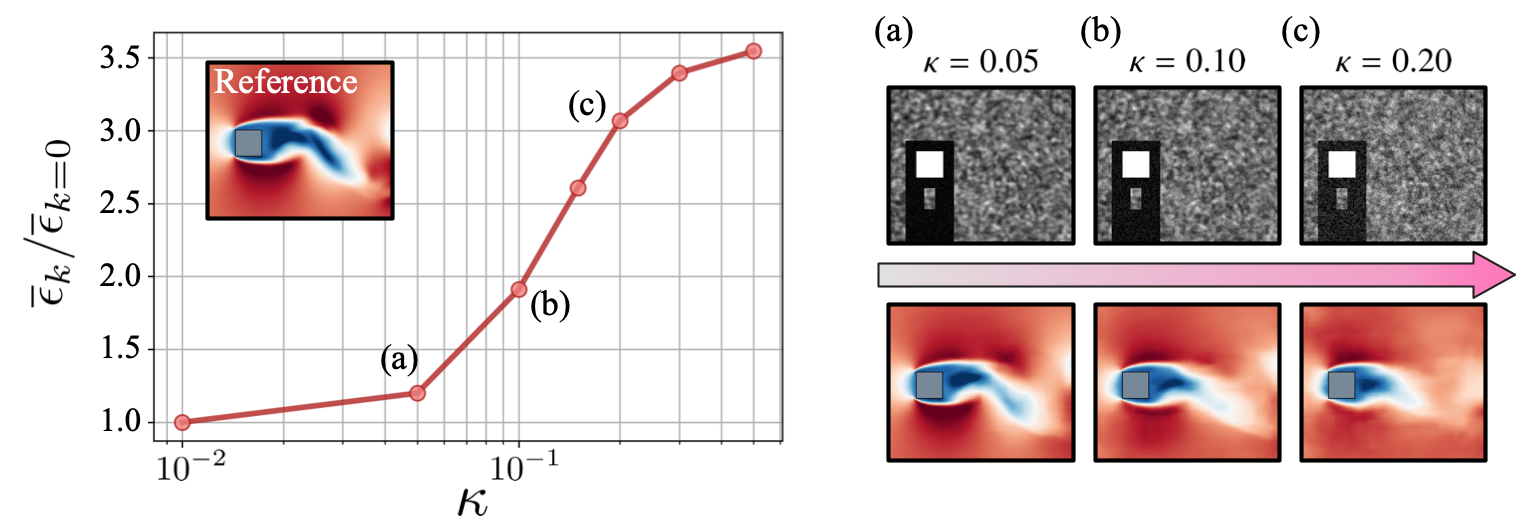}
    \caption{Robustness against noise of the proposed machine learned model for a square cylinder flow. \fg{Example APIs with the noise magnitude of (a) $\kappa=0.05$, (b) $\kappa=0.10$, and (c) $\kappa=0.20$ are shown with the estimated fields.}}
    \label{fig:noise-robustness}
\end{figure}

We then evaluate robustness against a noisy input ${\bm q}_{\rm noise}={\bm q}+\kappa {\bm n}$, where $\bm n$ denotes the Gaussian noise and $\kappa$ is its amplitude, as presented in Fig.~\ref{fig:noise-robustness}.
This assessment can be one of the benchmarks for applications to experiments.
Note here that the scale of the vertical axis in Fig.~\ref{fig:noise-robustness} is normalized by the error in the case without noise.
Up to $\kappa=0.10$, the flow fields show no significant difference against the reference field presented on the upper-left portion of the error plot.
However, with $\kappa=0.20$, the finer scale motion in the wake region cannot be estimated, although the large scale motion around the bluff body is still captured.

\begin{figure}
    \centering
    \includegraphics[width=\textwidth]{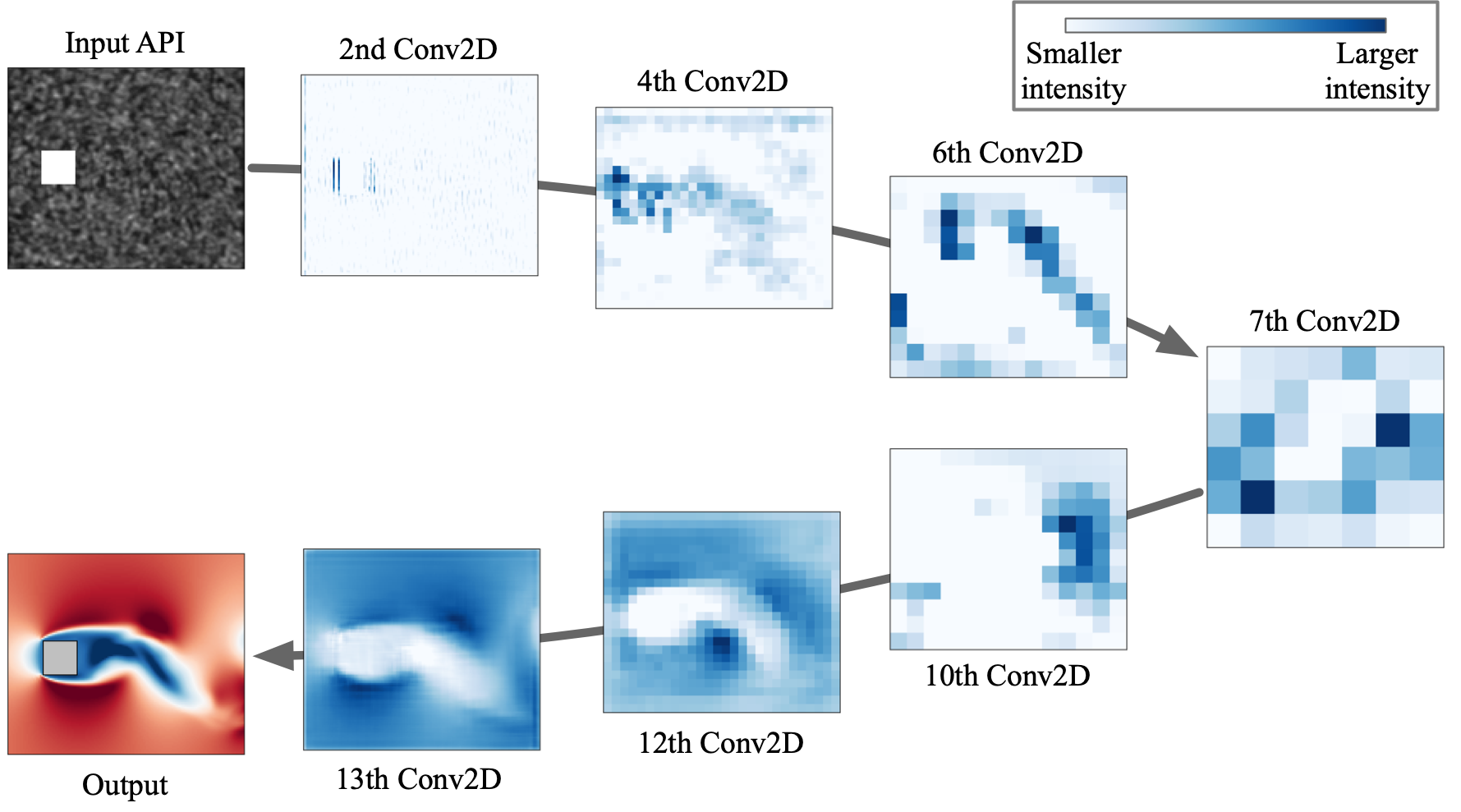}
    \caption{Output of each hidden layer with an example of a single square cylinder flow.}
    \label{fig:vis_hidden-layer}
\end{figure}
\begin{figure}
    \centering
    \includegraphics[width=0.9\textwidth]{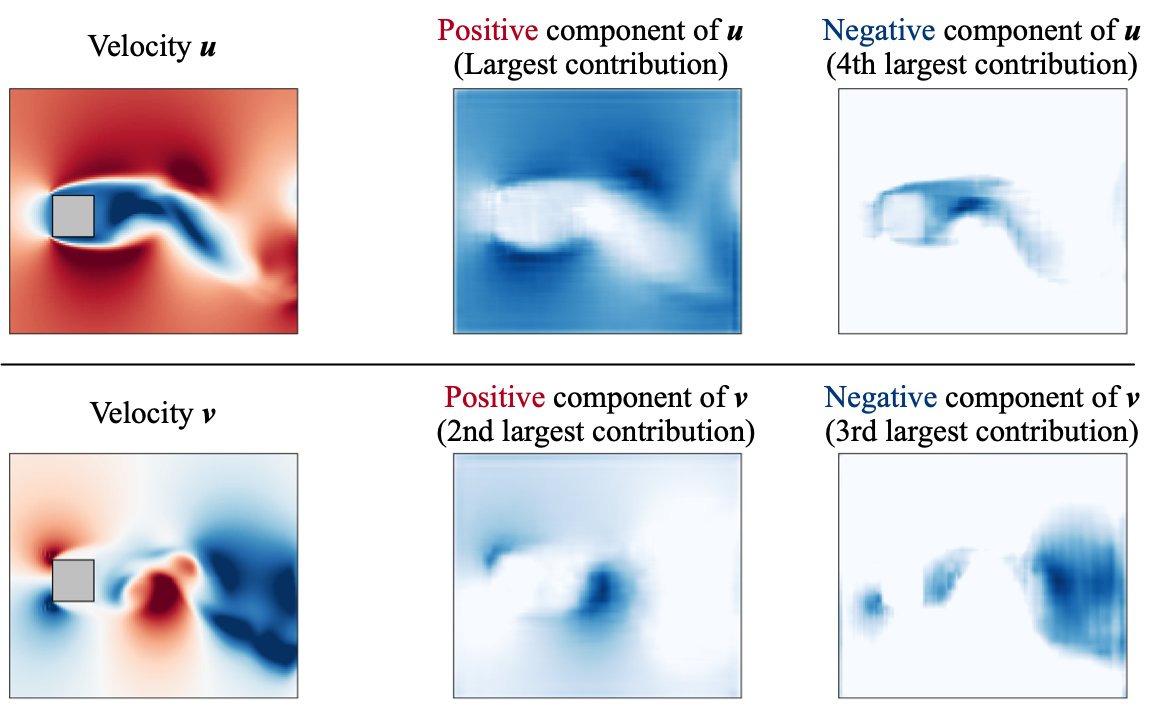}
    \caption{Visualization at the 14th linear layer of machine learned model with a single square cylinder. Orders above each figure indicate {their} contribution{s} to {the} output.}
    \label{fig:last_layer}
\end{figure}

Before applying our machine learned model to experimental data, we investigate in which regions the machine learned model has strong interests for estimating the velocity fields.
Let us visualize the output of each hidden layer.
\fg{Note that hidden layers are the layers placed between the input and output layers of neural networks, i.e., layers other than `Input' and `14th Conv2D' layer in our model summarized in Table~\ref{tb:network_structure}.}
Example outputs of each layer are shown in Fig.~\ref{fig:vis_hidden-layer}.
Since each hidden layer has 16 or 32 outputs according to the number of filters, the output data with the largest average intensity is shown as the representatives.
Note that the outputs of each hidden layer have only positive values because the rectified linear unit (ReLU) function is utilized as the activation function in the present study, and therefore, larger values indicate stronger interests.
The outputs of 2nd and 4th layer clearly show that the machine learned model has a strong interest on the location of the square cylinder.
In contrast, for the layers near the output, e.g. 12th and 13th layers, the machine learned model focuses on the wake region.
This observation implies that the machine learned model first finds out the location of the square cylinder and then focuses on the regions of velocity fluctuations so as to output appropriate velocity fields.
We are also able to see clearly the second trend from the output of 13th layer, as summarized in Fig.~\ref{fig:last_layer}.
We show here the four most dominant outputs.
A noteworthy observation is that the positive values and negative values of velocity field are separated in different channels.
This is likely because the machine learned model needs to represent negative value of velocity field at 14th layer, since we use ReLU
{---} which has only positive range {---} 
as the activation function in the upstream layers.
Although we visualized the filters in convolutional neural network (CNN) such as Figs.~\ref{fig:vis_hidden-layer} and~\ref{fig:last_layer}, different tools, e.g., Grad-CAM~\cite{selvaraju2016grad,selvaraju2017grad,JZV2020}, can also be considered to interpret the results obtained by machine learning models.

\begin{figure}
    \centering
    \includegraphics[width=\textwidth]{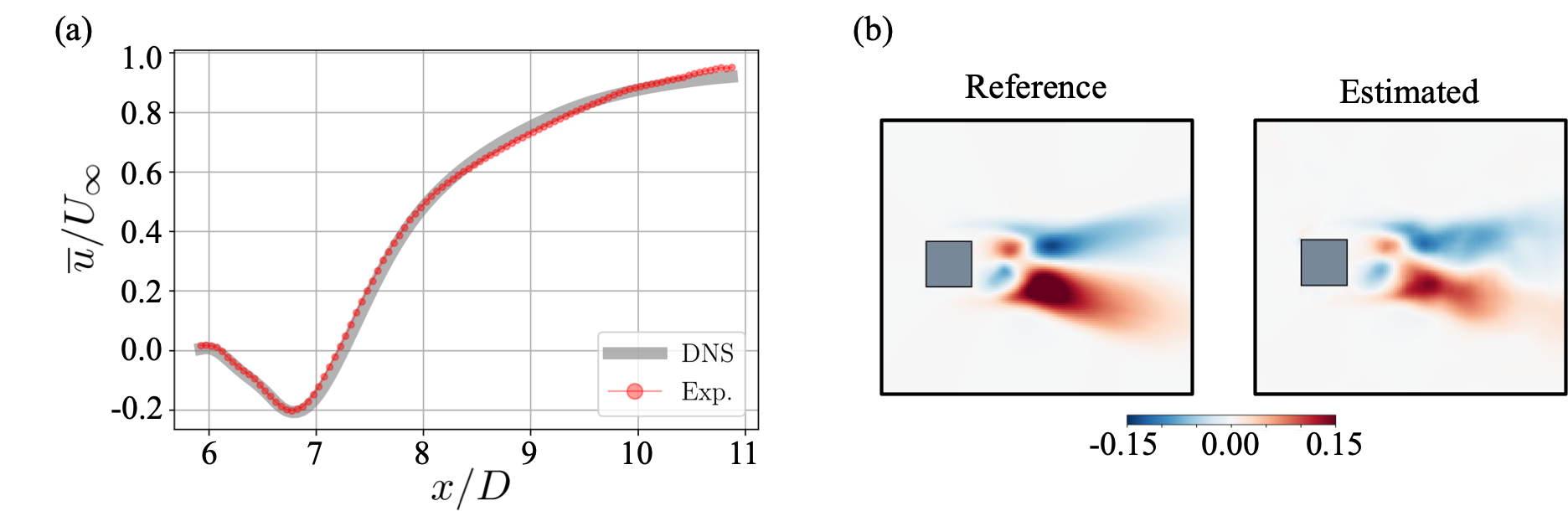}
    \caption{Application of the machine learned model to experimental particle images. (a) Mean centerline velocity; (b) Reynolds shear stress $\overline{u^\prime v^\prime}$. `Exp.' indicates the result of machine learned model with experimental images as the input.}
    \label{fig:1SC_umean-RS}
\end{figure}

We then apply the machine learned model to {estimate the velocity field from} our experimental images.
To evaluate the result, the centerline velocity and distribution of the Reynolds shear stress $\overline{u^\prime v^\prime}$ are assessed in Fig.~\ref{fig:1SC_umean-RS}, since the correct answers corresponding to the instantaneous experimental images do not exist.
As shown in Fig.~\ref{fig:1SC_umean-RS}(a), the mean velocity profile of the estimated fields is in excellent agreement with the reference DNS data.
For the Reynolds shear stress in Fig.~\ref{fig:1SC_umean-RS}(b), the structure can be captured well, which confirms that the fluctuation components can also be estimated successfully, although the absolute value of the estimated field is slightly lower than that of DNS.

\begin{figure}
    \centering
    \includegraphics[width=\textwidth]{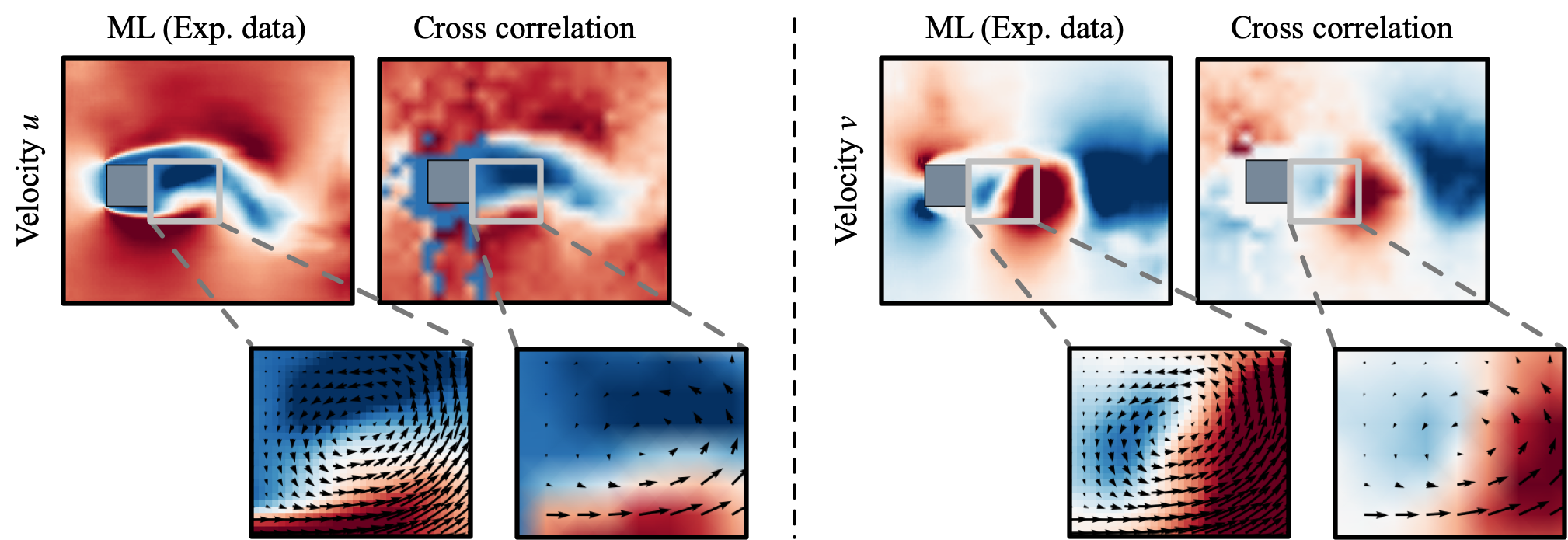}
    \caption{Comparison of velocity fields between the present model and the cross-correlation method.}
    \label{fig:PIV-vs-ML_1SC}
\end{figure}

An instantaneous flow field estimated using the machine learned model is also compared to the conventional cross-correlation method~\cite{adrian2005}, as shown in Fig.~\ref{fig:PIV-vs-ML_1SC}.
The flow field can be recovered successfully by using the machine learned model.
On the other hand, the result with the cross-correlation method is grossly affected by the missing regions.
It suggests that the present machine learned model can retain the reconstruction ability by explicitly giving the lacked portion to training data.
What is also noteworthy here is the density of velocity vectors obtained with the machine learning based estimation compared to the conventional cross-correlation method.
The machine learning based model can provide much finer resolution than the cross-correlation method, which can provide only one velocity vector from an interrogation window including some particles. 
Because the size of interrogation window is determined based on the experimental setup, such as the particle diameter, the particle number density, and the spatial resolution of the camera, it is usually tougher to obtain finer structures of flow field than a numerical simulation.
In contrast, since the machine learning model is trained to estimate vectors on every single computational grid points, it is able to estimate finer structures of the field~\cite{CZXG2019}.
Particularly in our cases, the machine learned model can provide approximately 40 folds denser field than the cross-correlation method.
This advantage enables us to find small-scale structures which cannot be captured with the conventional methods.
Note that, although there is a room for the parameters of PIV to be optimized to improve the quality of cross-correlation based velocity estimation in Fig.~\ref{fig:PIV-vs-ML_1SC}, the missing regions and the denser field cannot be obtained even if the optimization is performed.
\fg{Moreover, regarding the post-processing time, it takes approximately 40 minutes with CPU (Intel Core i7-8550U, 1.85 GHz) to perform cross correlation method for 10,000 snapshots.
In contrast, our machine learned model can handle the same amount of the snapshots approximately in 10 seconds with NVIDIA TESLA V100 graphics processing unit (GPU).
Furthermore, because there is no process to eliminate incorrect vectors or embed missing collocation points with 
neighboring 
vector information with machine learning, an overall processing time of ML-based estimator is expected to be significantly shorter than that of the traditional methods.
}

\subsection{Example 2: Wakes around two staggered square cylinders}

The problem setting of example 1 can be regarded as a relatively easy task for the machine learning model to augment the velocity field of a lacked region because the flow there was nearly steady.
For further investigation on the capability of the machine learning model, let us consider a flow around two staggered square cylinders~\cite{MGY2016}.

\begin{figure}
    \centering
    \includegraphics[width=\textwidth]{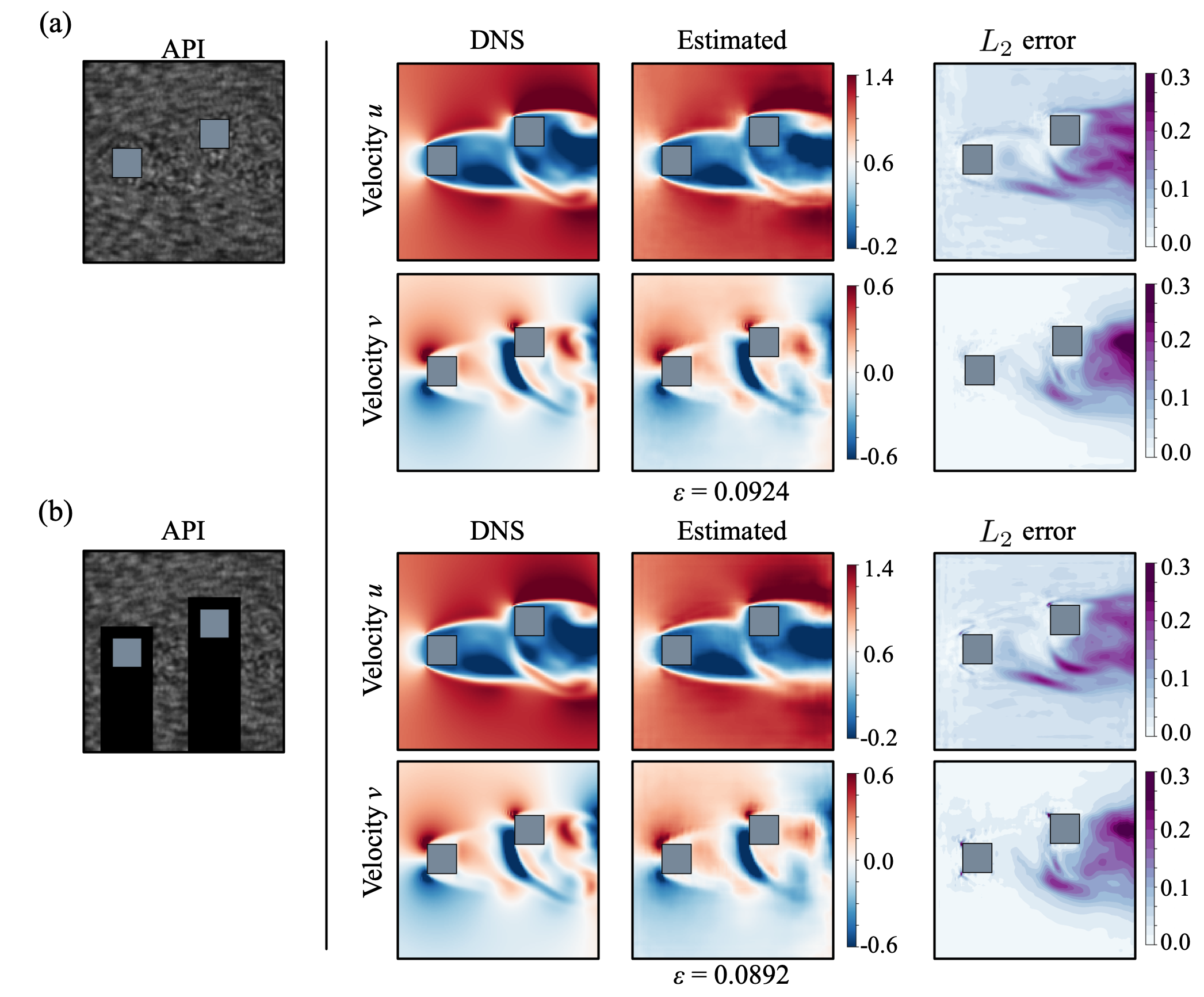}
    \caption{Velocity fields estimated from APIs of flow around two staggered square cylinders. : (a) full APIs $\bm{q}$; (b) lacked APIs $\widehat{\bm{q}}$.}
    \label{fig:2SC_API}
\end{figure}
\begin{figure}
    \centering
    \includegraphics[width=0.9\textwidth]{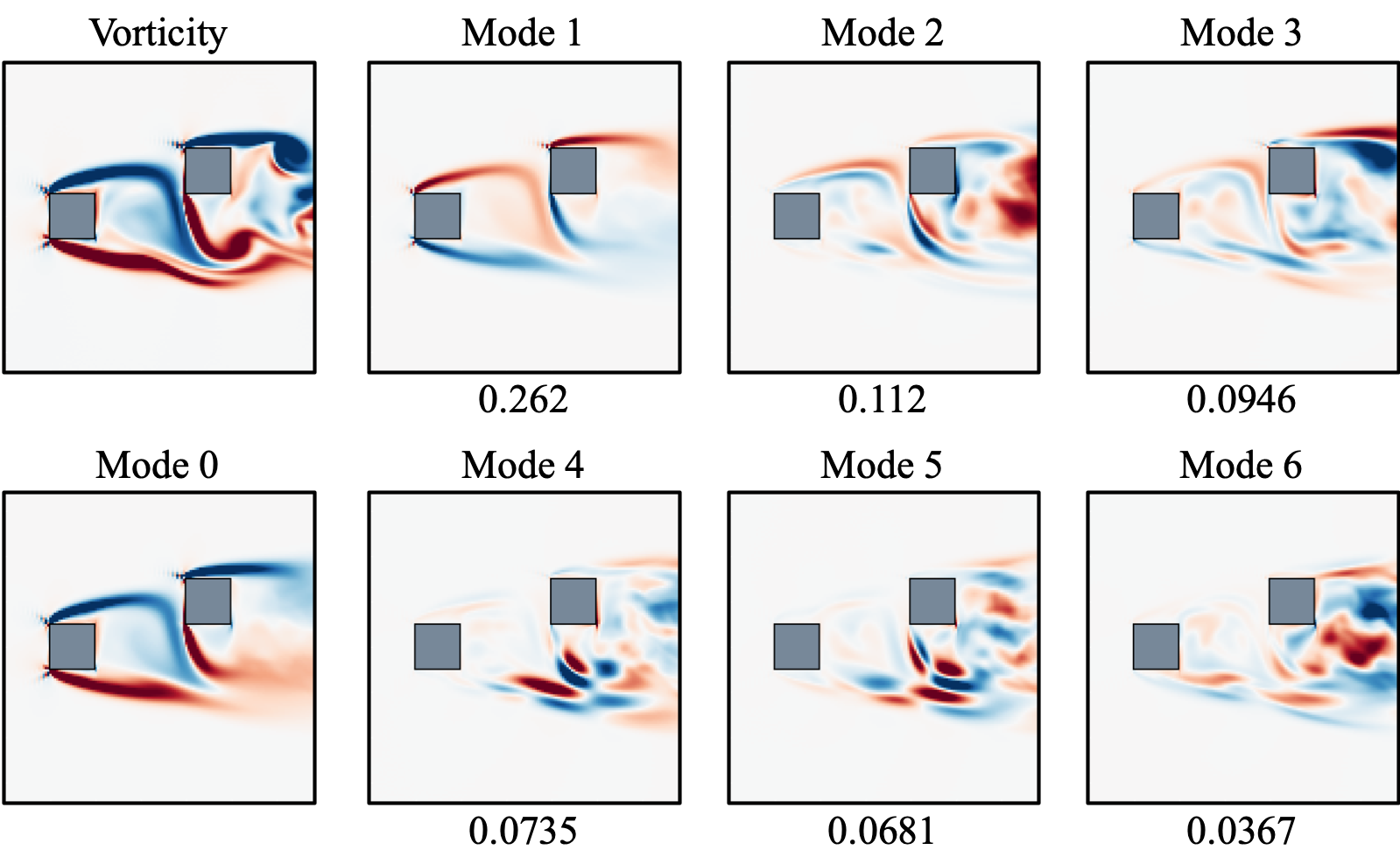}
    \caption{Six dominant POD modes of a vorticity field around two staggered square cylinders. Values listed below of each modes show contained energy rate of each mode.}
    \label{fig:2SC_POD}
\end{figure}

The flow fields estimated from APIs are shown in Fig.~\ref{fig:2SC_API}.
With both full and lacked inputs, the estimated flow fields are in reasonable agreement with the reference DNS data.
The $L_2$ error norms listed underneath the figures show approximately $10\%$, which are almost the same as that of example 1.
However, we should note that the error is concentrated in the region below the second square cylinder, where the complex structure is hidden as shown in Fig.~\ref{fig:2SC_API}.
It can also be seen from the proper orthogonal decomposition (POD) analysis with vorticity field obtained by DNS as assessed in Fig.~\ref{fig:2SC_POD}.
It is striking that the complex structures below the second square cylinder can be seen on POD modes 4 and 5, a sum of which contains approximately $15\%$ of the kinetic energy.
Since there is no significant difference in the $L_2$ error norm between full and lacked APIs, this POD analysis tells us the fact that the machine learned model for the lacked APIs $\widehat{\cal{F}}$ was able to estimate these complex structures as well as the model for full APIs $\cal{F}$ even the structures are masked in advance.

\begin{figure}
    \centering
    \includegraphics[width=\textwidth]{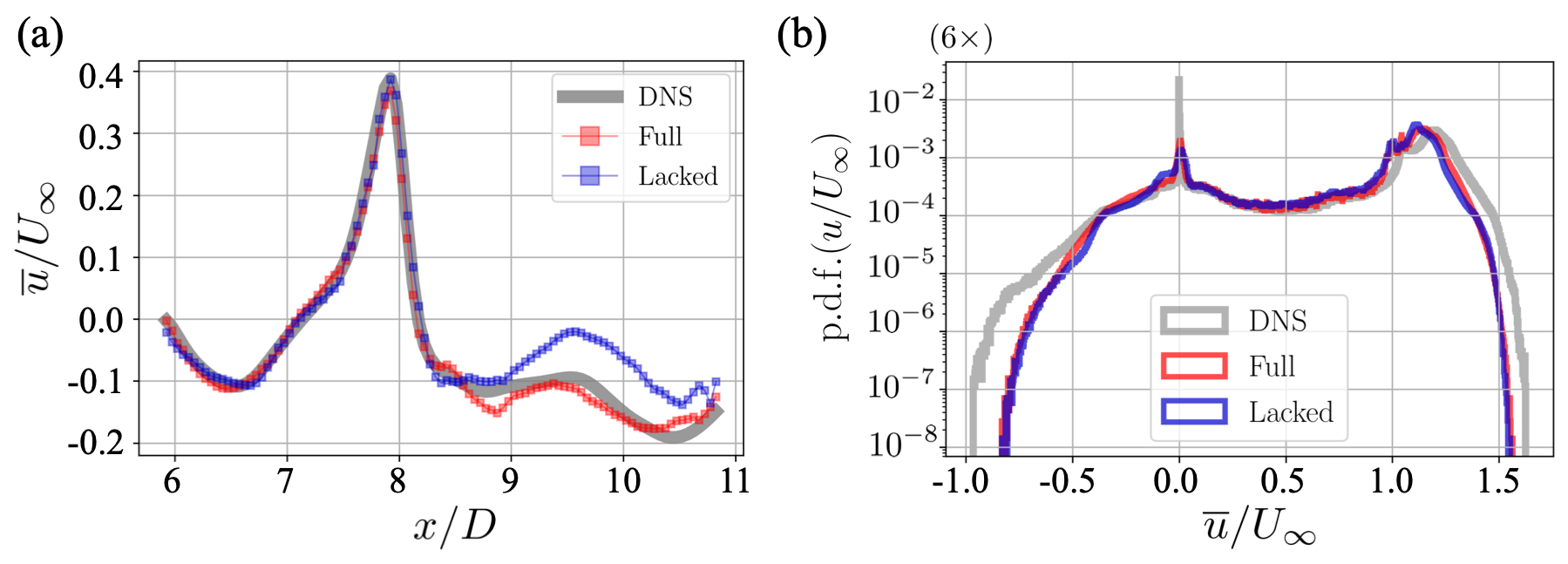}
    \caption{(a) Mean centerline velocity and (b) probability density function ({PDF}) of streamwise velocity estimated from full and lacked APIs of flow around two staggered square cylinders.}
    \label{fig:2SC_API_umean_pdf}
\end{figure}

For further assessment, the mean centerline velocity profile and the PDF are presented in Fig.~\ref{fig:2SC_API_umean_pdf}.
The overall profile of the mean centerline velocity agrees well with the DNS data, although the slight mismatch can be seen on the vortex shedding region.
The reason for this mismatch can be also seen in the PDF, which suggests that the machine learned model fails to estimate the low probability events.
This is, again, likely because the training process is designed to minimize the $L_2$ error, to which lower probability events have less contributions.

\begin{figure}
    \centering
    \includegraphics[width=\textwidth]{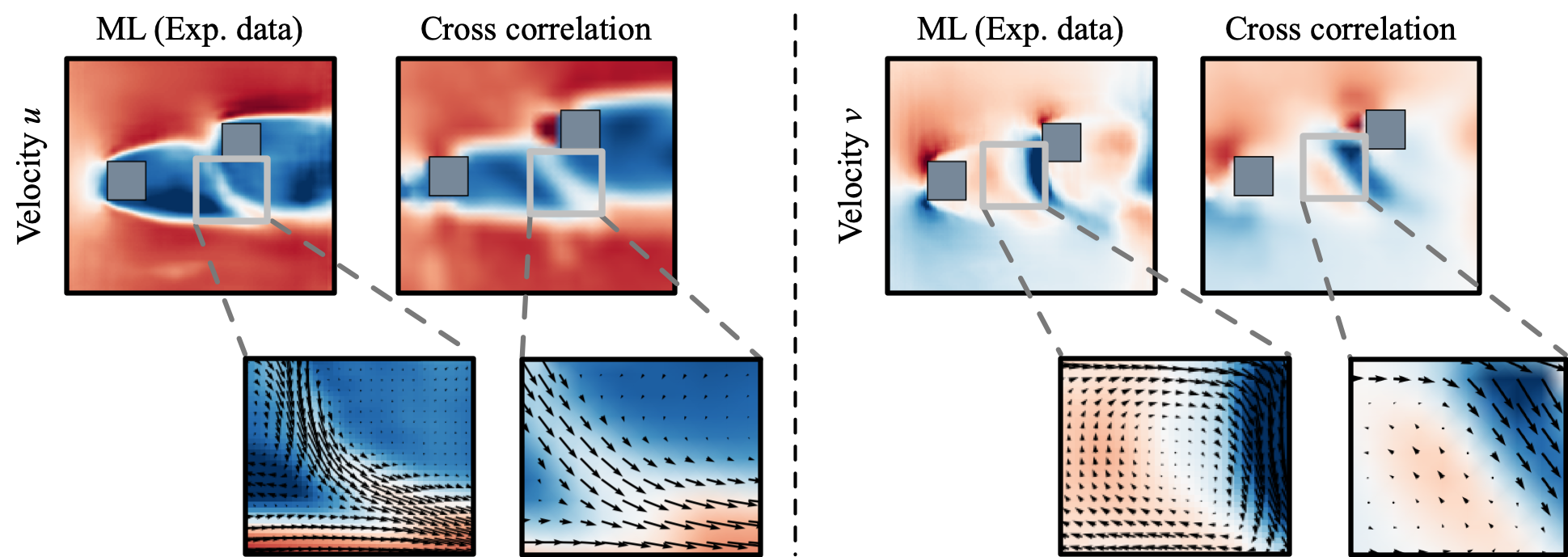}
    \caption{Comparison of velocity fields around two staggered square cylinders between the present model and the cross-correlation method.}
    \label{fig:2SC_API_PIV_VS_ML}
\end{figure}

The estimated flow fields with the experimental image inputs of a flow around two staggered square cylinders are presented in Fig.~\ref{fig:2SC_API_PIV_VS_ML}.
Similar to the single cylinder example, denser field can be seen with the use of the machine learned model.
It is striking that the differences in the velocity magnitude are clearer than that with the conventional cross-correlation due to the finer structures.
Although we have no correct answer for this experiment, the results suggest again that the machine learned model enables us to extract the hidden structures of flow fields.

\begin{figure}
    \centering
    \includegraphics[width=\textwidth]{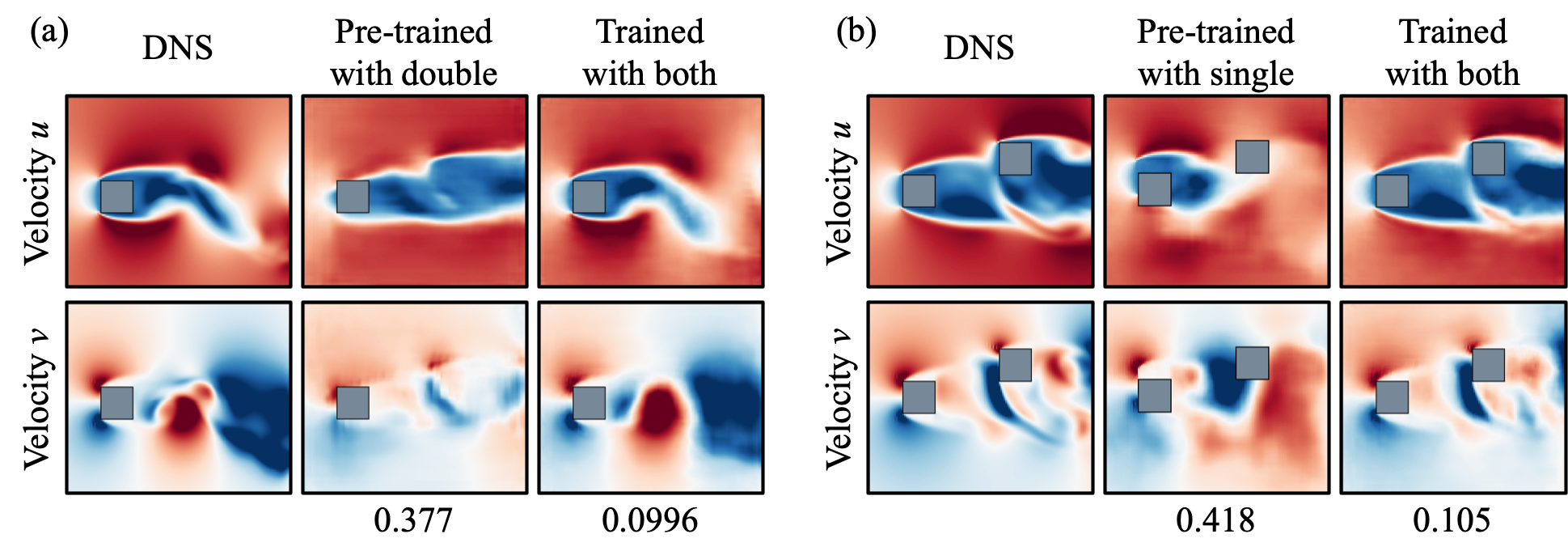}
    \caption{Estimation of flows around (a) a square cylinder and (b) two staggered square cylinders by a pre-trained model with the other field and a model trained with both fields.}
    \label{fig:unlearned_data}
\end{figure}

Finally, we perform three kinds of tests to investigate the influence of the alignment of bluff bodies on the {machine learning} based estimation as follows:
\begin{enumerate}
    \item Apply the machine learned model pre-trained with flows around two staggered square cylinders to a flow around a single square cylinder.
    \item Apply the machine learned model pre-trained with flows around a single square cylinder to a flow around two staggered square cylinders.
    \item Apply the machine learned model pre-trained with both flows to both test data.
\end{enumerate}
Through these tests, we can see whether the present model can be utilized to a flow around unseen bluff body alignments or not.
Figure~\ref{fig:unlearned_data} summarizes the results of these tests.
For Fig.~\ref{fig:unlearned_data}(a), APIs of flow around a square cylinder are used as the test data.
The model pre-trained with two square cylinders fails to estimate the correct field ---
the estimation is highly affected by the alignment of square cylinders of training data.
The same trends can be seen in the case of estimating a flow around two square cylinders by using a model trained with a single square cylinder, as shown in Fig.~\ref{fig:unlearned_data}(b).
Since the machine learned model trained with a single flow configuration has relatively low applicability to flows around unseen bodies, we further assess the ability of the model trained using  both types of flow fields as stated as procedure 3 above.
As presented in Fig.~\ref{fig:unlearned_data}, both estimated flow fields are in nice agreement with the reference DNS data.
The $L_2$ error norm shown underneath each estimated field, is approximately $10\%$, which is almost the same as those in the previous examples presented in Figs.~\ref{fig:1SC_API} and~\ref{fig:2SC_API}. 
These results indicate that we need to be careful in preparing the training data for machine learning based particle image velocimetry (PIV) data augmentation, especially in the case of estimation on flows around bluff bodies.

\begin{figure}
    \centering
    \includegraphics[width=\textwidth]{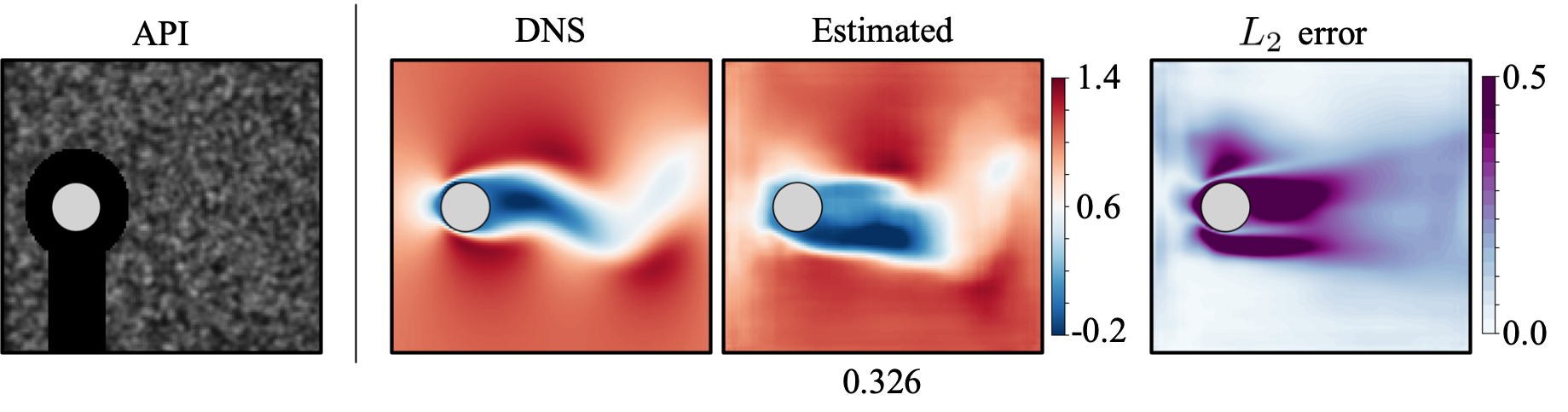}
    \caption{Estimation of a flow around circular cylinder at ${\rm Re}_D=100$ using the model pre-trained with both single and double square cylinders.}
    \label{fig:circular_cylinder}
\end{figure}

Furthermore, to check the applicability to flows around other bluff body shapes, we use the trained machine learning model to a wake of circular cylinder wake at ${\rm Re}_D=100$.
The machine learned model is pre-trained with both single and double square cylinders, which was also used for the assessment in Fig.~\ref{fig:unlearned_data}.
The cylinder wake data is prepared by DNS and a no-slip boundary condition on the cylinder surface is provided using an immersed boundary method~\cite{kor2017} as detailed in Appendix B.
Since both the bluff body shape and the Reynolds number differ from those used for training, this is a challenging task to estimate the velocity field properly.
As shown in Fig.~\ref{fig:circular_cylinder}, the machine learned model cannot estimate the velocity field properly, mainly near the region around the body.
This result also motivates us to care for the proper choice of training data in estimating the flows around a bluff body with a supervised manner.

\section{Conclusions}
\label{sec:conclusion}

We presented a machine learning based data estimation method for particle image velocimetry (PIV) considering a typical experimental limitations, i.e., missing regions.
The training data set was prepared using a direct numerical simulation (DNS) and artificial particle images (APIs).
Two types of data, i.e., full APIs (no data missing region) and lacked APIs (with a fixed data missing region), were considered to examine the applicability of the machine learning model to experimental situations.
The lacked regions were set by referencing to our exact wind tunnel setup.
We utilized a model similar to a convolutional neural network based autoencoder as the machine learning model.

As the first example, we considered a flow around a single square cylinder at ${\rm Re}_D=300$.
The estimated flow fields with both full and lacked APIs showed great agreement with the reference DNS data.
It was found that the machine learned model can estimate the flow field from experimental images which has a lacked region, while the conventional cross correlation method is heavily affected by the lacked region.
These trends could be seen also in the statistical assessment using the mean centerline velocity profile and the Reynolds shear stress distribution.
In addition, by using the machine learning model, finer velocity fields could be obtained than that with a cross correlation method.

For further assessment on more complex flow fields, we applied the proposed method to a flow around two staggered square cylinders.
Despite that the flow field contains finer and complex structures, the machine learned model could estimate the velocity fields from APIs with $L_2$ error norm less than $10\%$.
We also found that even the complex structures, with the energy rate of $15\%$, are hidden by the alignment of two square cylinders, the machine learned model for lacked data is able to estimate the field as well as the model for full data.

We also investigated the applicability of pre-trained model to a flow around unlearned alignment and configuration of bluff bodies.
The results indicated that estimation with the machine learned model is affected by the alignment of bluff bodies.
We also found that the machine learned model could be able to establish the function in not only estimating the field, but also recognizing the alignment of bluff bodies if proper training data sets are given.
This point was further confirmed in the additional test using the flow around a circular cylinder at a different Reynolds number.

\fg{
For more practical uses, cares should be taken for the choice of training dataset.
Cai et al.~\cite{CZXG2019}, who have succeeded in estimating various flow fields with a single machine learning model, reported that a machine learning model can handle a wide range of experimental datasets by giving sufficient training data.
On the numerical side, several studies reported the possibility to generalize the models against the Reynolds number~\cite{HFMF2020b} and flow configuration~\cite{HFMF2020a,Gentech2020} from limited available training data.
The combination with these techniques can be promoted in the future.}

\fg{The use of unsupervised learning can also be one of the considerable approaches toward a construction of generalizable machine learning model.
Kim et al.~\cite{GANSRkim2021} utilized a GAN-based unsupervised network for a super-resolution task of turbulent flow and demonstrated its universality to higher Reynolds number flows.
In experiment, Zhang and Piggott~\cite{ZP2020} have recently demonstrated the possibility of the use of unsupervised machine learning for PIV velocity estimation by unifying with classic optical flow methods, 
and Lagemann et al.~\cite{LLMS2021} have applied the recurrent all-pairs field transforms (RAFT) to the optical flow method to achieve a higher accuracy and generality.}
We can also expect the fusion of the knowledge obtained through our presentation and these new trials, which enables us to believe that our approach can be one of bridges toward practical and generalized applications of machine learning for PIV situation.

\fg{Moreover, the proposed method can be extended to handle volumetric PIV and time-resolved PIV by replacing the two-dimensional (2D)-CNN to a three-dimensional (3D)-CNN, which is capable of handling volumetric or spatio-temporal flow data\cite{nakamura2020extension,FFT2020b}.
Moreover, by combining 2D- and 3D-CNN, for example, it is also able to reconstruct volumetric velocity data from several sectional experimental images~\cite{matsuo2021supervised}.

Summarizing above, we believe that the present concept, which has the capability to complement the defects of the conventional cross correlation method, can be applied to more complex flows and data form which contain various scales of structures.}

\section*{Acknowledgements}
This work was supported through JSPS KAKENHI (Grant Number 18H03758\fg{, 21H05007}) by Japan Society for the Promotion of Science.
The authors acknowledge Mr. Takaaki Murata, Mr. Taichi Nakamura (Keio University) and Mr. Kazuto Hasegawa (Keio University, Polimi) for fruitful discussions.
The authors also thank Mr. Hikaru Murakami (Keio University) for sharing his DNS data.

\section*{Appendix A: Details of PIV setup}
\begin{figure}
    \centering
    \includegraphics[width=0.45\textwidth]{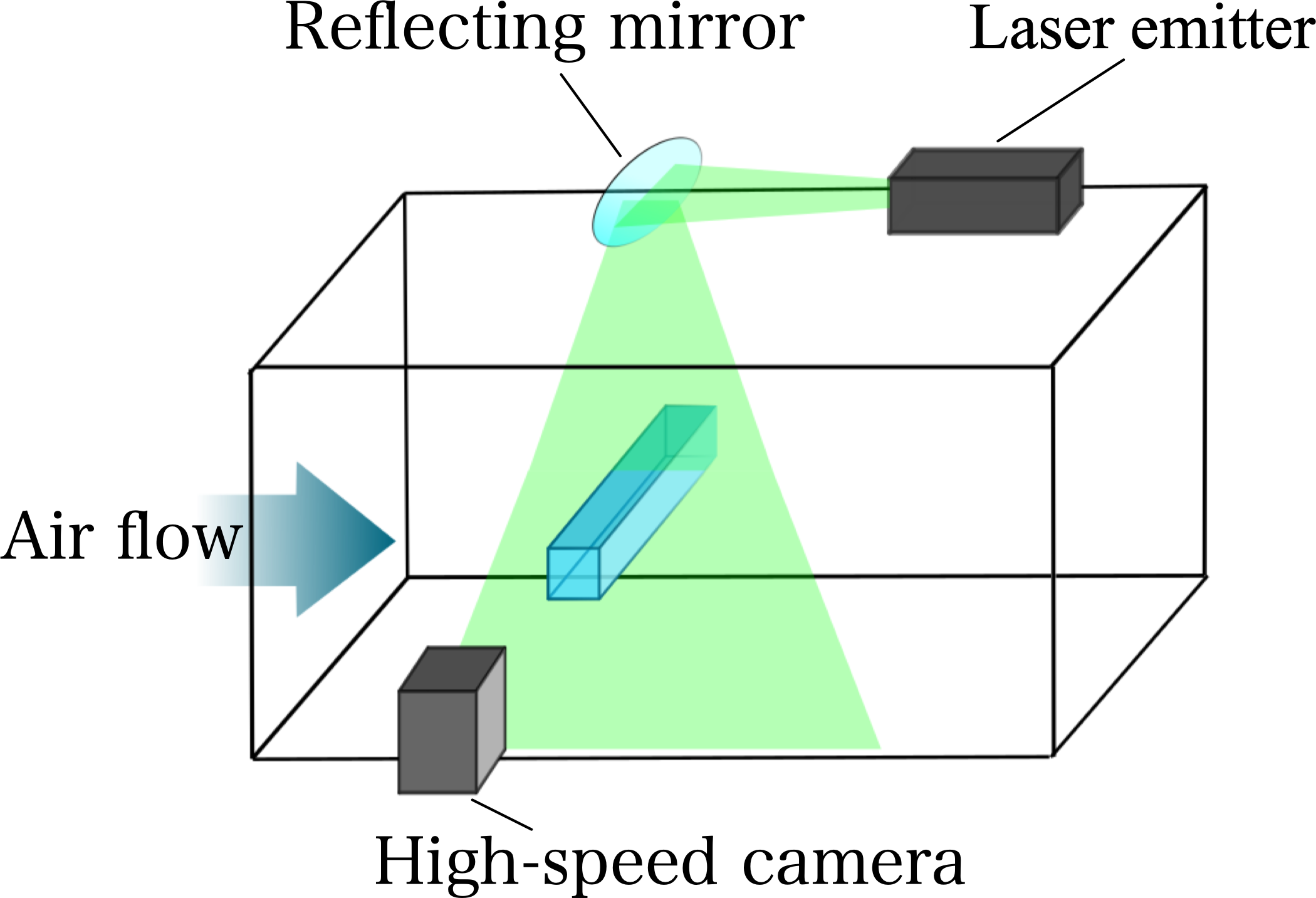}
    \caption{Illustration of our PIV setup utilized in this study.}
    \label{fig:PIV_setup}
\end{figure}

Here, we introduce the fundamental setup of particle image velocimetry (PIV) utilized in this study.
Let us present in Fig. \ref{fig:PIV_setup} the illustration of our setup.
We use a compact closed-loop wind tunnel (SGD-150, San Technologies, Co.~Ltd., Japan)
whose test section has a size of $300\ \si{mm}\times150\ \si{mm}\times 150\ \si{mm}$.
In the case of a single square cylinder, the cylinder is located at $94.0\ \si{mm}$ from the inlet section and $71.5\ \si{mm}$ from the bottom of the wind tunnel.
In the case of two staggered square cylinders, an additional cylinder is placed at $9\ \si{mm}$ and $6\ \si{mm}$ away in $x$ and $y$ directions, respectively, from the upstream cylinder.
To match the Reynolds number to that of DNS, one side of a square cylinder is set to $3\ \si{mm}$ and the flow speed is set to $1.5\ \si{m/s}$.

We use a high-speed camera (K5-USB monochrome 8GB, Kato Koken, Co.~Ltd., Japan) with a Nikkor $85\ \si{mm}$ F/1.6D (Nikon Corp., Japan) lens to capture the particle images of flow fields.
The resolution of the camera is $640 \times 480$ pixels, and the image depth is 12 bits.
The frame rate is set to 4000 frame/s.
As an illuminating equipment, we use a laser-diode-excited YVO4 solid-state continuous-wave laser (G1000, Kato Koken) with the wavelength of $532\ \si{nm}$ and the sheet thickness of $2\ \si{mm}$.
The excited laser is reflected by a mirror of $140\ \si{mm}\times 100\ \si{mm}$ (MRGA2H3, Misumi Group Inc., Japan).
For seeding, we use a seeding generator (CTS-1000, Seika Digital Image Corp., Japan) to seed nominally $2.0\ \si{\micro m}$ dioctyl sebacate (DOS) droplets by pumping compressed air of approximately 5 atoms into the seeding generator.
A Windows 64-bit PC is utilized for operating the high-speed camera (k-Software, Kato Koken) and performing PIV (FlowExpert2D2C, Kato Koken).
The height and width of interrogation window is set to 32 pixels considering the particle image diameter and by monitoring the cross correlation.
The actual size of a single pixel is $3.75\times10^{-2}\ \si{mm}$.

\section*{Appendix B: Details of DNS and data preparation}
\begin{figure}
    \centering
    \includegraphics[width=\textwidth]{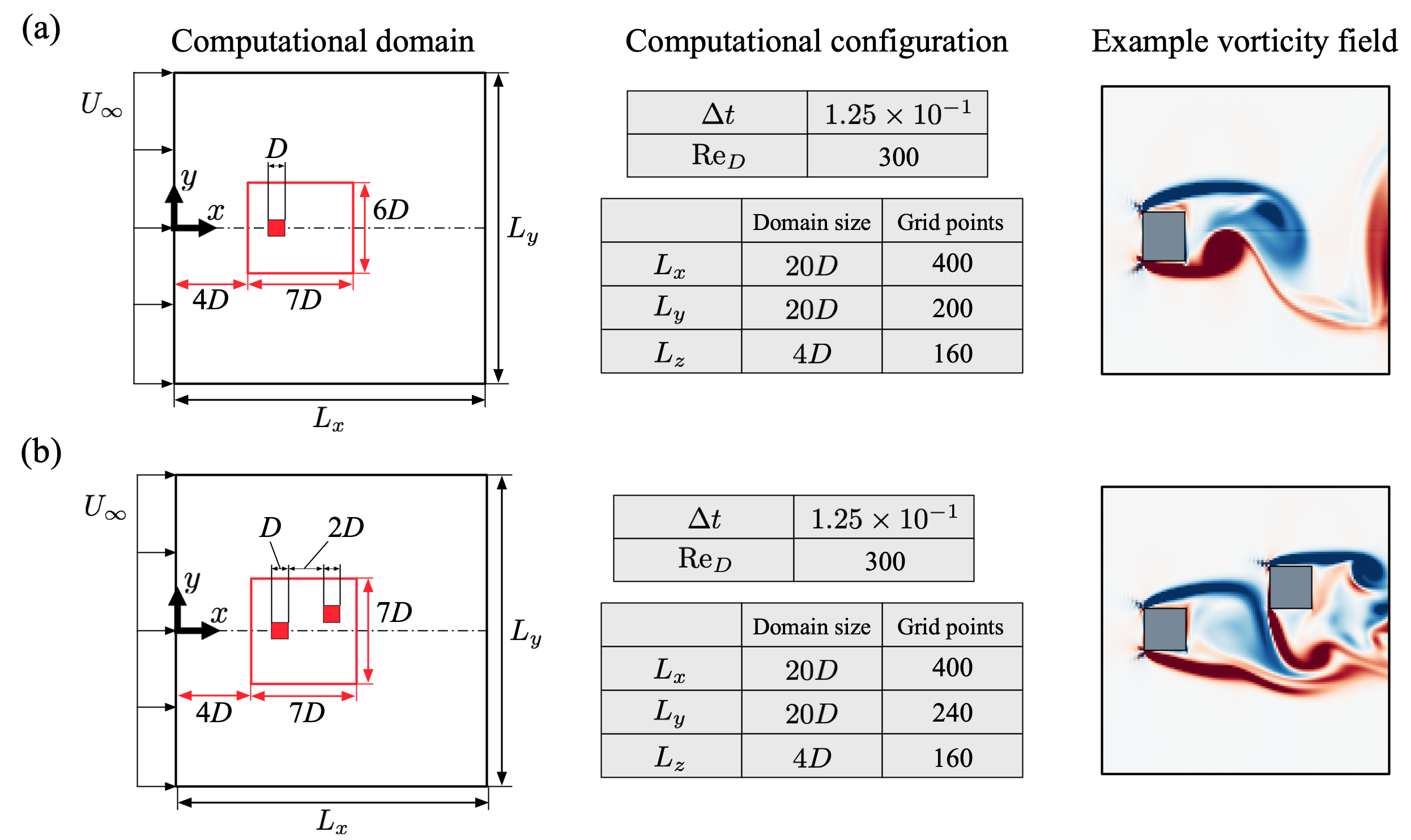}
    \caption{Computational domain and vorticity field of flow around (a) a square cylinder and (b) two staggered square cylinders.}
    \label{fig:comp-domain}
\end{figure}
\begin{figure}
    \centering
    \includegraphics[width=\textwidth]{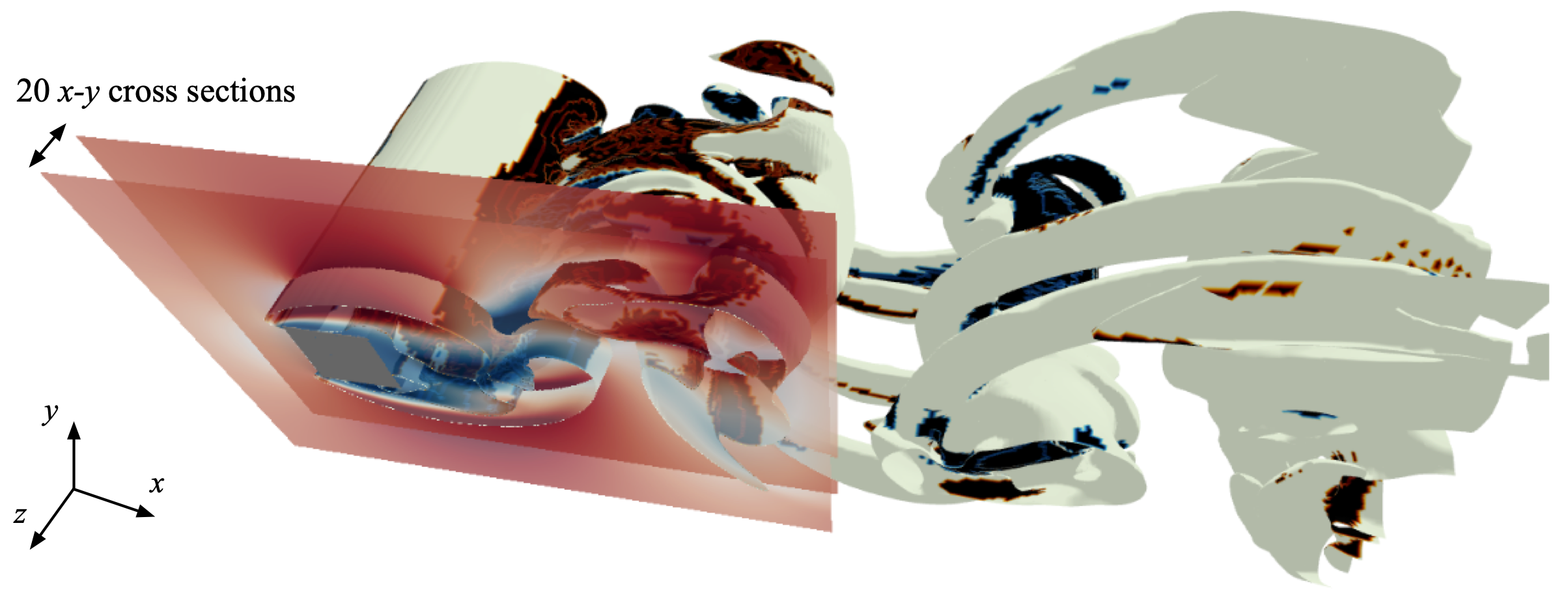}
    \caption{Preparation for training data in the present study. Twenty $x-y$ sections of three-dimensional data with $500$ time steps are extracted.}
    \label{fig:DNS_extraction}
\end{figure}

Here, the detailed information of direct numerical simulation (DNS) of covered flows are provided.
We have mainly considered a flow around a square cylinder at the Reynolds number ${\rm Re}_D=300$ based on the free-stream velocity and the side length of the cylinder $D$.
The training data set is prepared by a direct numerical simulation (DNS) with numerically solving of the incompressible Navier--Stokes equations with a penalization term~\cite{volumePenal1994},
\begin{eqnarray}
        {\bm{\nabla}} \cdot {\bm v}&=&0 \\
        \frac{\partial {\bm v}}{\partial t} + {\bm{\nabla}} \cdot \left({\bm{v}}{\bm{v}}\right)&=&-{\bm{\nabla}} p + \frac{1}{{\rm Re}_D}{\bm{\nabla}}^2\bm{v}+\lambda \chi\left({\bm v}_b-{\bm v}\right)\\ \nonumber
\end{eqnarray}
where ${\bm v}=\{u,v,w\}$, $p$ and ${\rm Re}_D$ are the non-dimensionalized velocity vector, pressure and Reynolds number respectively.
\fg{All quantities are non-dimensionalized by 
the fluid density $\rho$,
the side length of the square cylinder $D$, and the free-stream velocity $U_{\infty}$.}
The penalization term represents an object with a penalty parameter $\lambda$, a mask value $\chi$, and a velocity vector of a flow inside the object ${\bm v}_b$ which is $0$.
The mask value $\chi$ takes $0$ outside of an object and $1$ inside of an object.

As shown in Fig.~\ref{fig:comp-domain}(a), the size of the computational domain here is $\left(L_x \times L_y \times L_z\right)=\left(20D\times20D\times4D\right)$.
{The computational time step is $\Delta t=5.0\times10^{-2}$, which results in the maximum Courant number around $0.3$.}
Our DNS code has been verified with Franke et al.~\cite{FRS1990} and Robichaux et al.~\cite{RBV1999}.

For the training data of the first case with a single square cylinder, we focus on the volume around the square cylinder such that $\left(7D\times6D\times0.5D\right)$ shown by red lines in Fig.~\ref{fig:comp-domain}(a).
The number of grid points of the extracted region is $(N_x^\sharp\times N_y^\sharp\times N_z^\sharp)=(140\times120\times20)$.
Since three-dimensional vortex structure is present at ${\rm Re}_D=300$~\cite{BA2018}, twenty $x-y$ cross sections in $z$ direction are used for the training data to account for the three-dimensionality.
For the training data we use 10000 $x-y$ cross sectional snapshots, which include 500 time steps per twenty positions in $z$ direction, as illustrated in Fig.~\ref{fig:DNS_extraction}.
For the test data, we extract 160 $x-y$ cross sections from five different instantaneous three-dimensional snapshots ; namely, 800 snapshots are utilized for evaluation.
Note that each instantaneous field is chosen to be distanced from each other, which enables us to evaluate the estimation ability by various states of the flow.

The training data for the two staggered square cylinders are prepared similarly to the first example. 
The computational domain with the size and grid points are summarized in Fig.~\ref{fig:comp-domain}(b).
The Reynolds number is set to ${\rm Re}_D=300$ and the time step for the DNS is $\Delta t=0.125$.
The flow includes more complex structure, especially below the second square cylinder as shown in the vorticity map of Fig.~\ref{fig:comp-domain}(b).

The test data for a flow around a circular cylinder at ${\rm Re}_D=100$ is also prepared by two-dimensional direct numerical simulation.
The non-slip boundary condition on the cylinder surface is imposed using an immersed boundary method~\cite{kor2017} instead of penalization.
The computational time step is $\Delta t=2.5\times10^{-3}$ and the domain size is $(L_x, L_y)=(25.6, 20.0)$, with the center of the cylinder located at $(x, y)=(9,0)$.
For the test data, we extract the region around a cylinder, $7.5 \leq x \leq 14.5$ and $-3.0 \leq y \leq 3.0$.
Although the number of grid points in the domain for the simulation is $(N_x^\sharp\times N_y^\sharp)=(280,240)$, reduction using a linear interpolation is performed to match the input image of the machine learned model pre-trained with square cylinders, i.e., $(\tilde{N^\sharp}_x, \tilde{N^\sharp}_y)=(140,120)$.

In the present study, normalization and standardization are not applied to both input and output attributes because we consider only flows around square and circular cylinders.
We have also confirmed that the magnitude scales of velocity fields have no significant influence in covered examples, i.e., single and two staggered square cylinders and circular cylinder, likely because the the order of magnitude is already unity for all quantities due to nondimensionalization.

%

\end{document}